\documentclass{jfm}

\usepackage{graphicx}
\usepackage{newtxtext}
\usepackage{newtxmath}
\usepackage{natbib}
\usepackage{hyperref}
\hypersetup{
    colorlinks = true,
    urlcolor   = blue,
    citecolor  = blue,
}

\usepackage{booktabs}
\usepackage{subfigure}
\usepackage[T1]{fontenc}
\usepackage{amsmath}

\usepackage{tikz}
\usetikzlibrary{3d,arrows.meta}

\newcommand{\dd}{\mathrm{d}}
\newcommand{\ii}{\mathrm{i}}
\newcommand{\ee}{\mathrm{e}}
\newcommand{\sgn}{\operatorname{sgn}}
\newcommand{\khat}{\hat{k}}

\title[Duct-mode radiation scattered by a nearby plate]{An LLM-assisted theoretical model for aeroengine installation noise scattering}

\author[Liang \& Huang]%
{ Sicong Liang,\ns
  Xun Huang  \thanks{Email address for correspondence:  huangxun@pku.edu.cn}
  }

\affiliation{ State Key Laboratory of Turbulence and Complex Systems, Aeronautics and Astronautics,
School of Mechanics and Engineering Science, Peking University, Beijing, China
}

\begin{document}
\sloppy
\maketitle

\begin{abstract}
A theoretical model is developed for the canonical problem of aeroengine fan noise impinging on, and scattering from, a nearby semi-infinite rigid plate. The model chains two Wiener--Hopf problems in different coordinate systems. The first stage solves the duct-radiation problem analytically in axisymmetric cylindrical coordinates. The second stage expands the field incident on the plate into a double angular spectrum of plane waves: the axial spectrum is inherited from the first stage, while the spanwise spectrum is generated by a Weyl--Sommerfeld expansion of the cylindrical harmonics. Each spectral component is then scattered by the plate.
The mathematical crux is the coordinate transformation between the axisymmetric duct frame and the Cartesian plate frame. The core physical insight--that the Weyl--Sommerfeld identity provides an exact spectral bridge between these systems--originates entirely with the authors. The heavy algebra of translating between the two spectral representations, together with the associated numerical verification, was carried out with the assistance of a large language model (LLM), under strict human oversight and independent numerical cross-checking. This division of labour enables the LLM to handle tedious symbolic manipulation without compromising rigour.
Results are presented for spinning modes radiating from a cylindrical duct with a subsonic jet into a quiescent ambient. The case study demonstrates that, under systematic verification, LLMs can accelerate tedious algebra in theoretical fluid mechanics, contributing a concrete empirical example to the broader discussion on AI's role in theoretical research.
\end{abstract}


\section{Introduction}
\label{sec:intro}

The Leiden Declaration \citep{leiden2026declaration} cautions against unreliable AI-generated proofs, particularly in safety-critical domains. Here we present a counterbalancing case study from classical fluid mechanics using Wiener--Hopf techniques \citep{noble1958,Huang:21nsr}. The essential physical insight originates entirely with the authors. We employ the large language model (LLM) solely for tedious symbolic manipulation and numerical verification (mainly in \S~\ref{sec:bridge}), under our strict oversight. This transparent collaboration -- where every derivation is independently validated -- demonstrates how and when LLMs can accelerate theoretical studies without compromising rigour, thereby offering a concrete empirical example from our work to the broader discussion on AI's role in applied mathematics and fluid dynamics.

In practice, the fan noise radiated from the open nozzle of an aeroengine bypass duct interacts with the wing, the pylon, or the fuselage, and the scattered field can differ markedly from the free radiation. A canonical building block of this installation effect is the scattering of duct radiation by a nearby rigid surface. The two separate modelling ingredients of the problem are classical in fluid mechanics and aeroacoustics. The radiation of a duct mode from the open end of a semi-infinite cylindrical pipe was solved by the Wiener--Hopf technique by \citet{munt1977,munt1990} for the hollow duct discharging a subsonic jet, extending the still-medium analysis of \citet{levine1948}; \citet{cargill1982} examined the low-frequency interplay between the unsteady jet flow and the radiated sound; \citet{rienstra1984} extended the solution to the annular bypass duct in a uniform subsonic flow; and a systematic numerical implementation, including efficient Cauchy-integral factorisations and contour deformations, was given by \citet{gabard2006}. Peake's group further carried the technique to coaxial multi-stream exhausts \citep{veitch2008}.

Next, the diffraction of a plane wave -- or a more general incident gust -- by the edge of a semi-infinite rigid plate is the canonical Wiener--Hopf problem \citep{noble1958,crighton1992}. Its aeroacoustic relevance was first established by the trailing-edge noise analysis of \citet{ffowcs1970}; \citet{rienstra1981} subsequently incorporated mean-flow effects into the diffraction solution. \citet{howe1991} introduced the first analytical model for serrated-edge noise, and \citet{ayton2013} developed high-frequency aerofoil-scattering theory. More recently, the matrix Wiener--Hopf formulation for spanwise-periodic serrated edges by \citet{huang2017} provides the direct theoretical foundation for the second stage of the present model.

Both lines have remained active. On the duct side, recent work includes thermoacoustic effects on spinning modes \citep{sugimoto2022}, slowly varying modes in sheared lined ducts \citep{rienstra2021}, a geometrical-theory-of-diffraction treatment of the duct termination benchmarked against the Wiener--Hopf solution \citep{jing2026}, and the extension to elliptic cross-sections \citep{wang2025}. On the plate side, \citet{baddoo2020} solved cascade scattering with complex boundary conditions, and \citet{hales2023} modelled the interaction of anisotropic turbulence with a rigid leading edge. 

On the applied side, installation noise -- the scattering of engine noise by the airframe -- has been computed numerically at ONERA with high-order computational aeroacoustics, with experimental validation for an over-wing nacelle configuration \citep{lorteau2019} and for a boundary-layer-ingesting rear-fuselage concept \citep{lorteau2022}, following the earlier  study of \citet{redonnet2010}; model-scale installed jet--pylon--wing experiments are reported by \citet{meloni2021}. Such full-geometry computations are expensive, and a theoretical model of the simplified duct-plate combination -- which captures the salient physical features -- remains a powerful tool for industrial applications while simultaneously posing a challenge to academia, thereby motivating this work.

Here the incident field on the plate is not a plane wave but the full, radially and azimuthally structured, partly evanescent near field of the duct exit. The purpose of the present work is to construct such a coupled model. The mathematical difficulty of the coupling is, at its core, a change of coordinate system: the duct's Wiener--Hopf solution was established in axisymmetric cylindrical coordinates and is naturally a one-dimensional axial spectrum of cylindrical harmonics, while the plate problem was developed in Cartesian coordinates and is naturally a two-dimensional spectrum of plane-wave gusts; a coupled model requires an exact translator between these two spectral and the associated coordinates. 
We achieve this by chaining the two Wiener--Hopf problems through an exact spectral bridge.

The first stage is left entirely untouched: the duct problem is solved once, in axisymmetric cylindrical coordinates, and is exploited only through its output, the half-range axial spectrum $F_+(u)$ of the vortex-sheet displacement. The bridge then re-expands the incident field on the plate plane into a double angular spectrum of plane waves: the axial dependence of each duct spectral component is already that of a convected gust, so the axial spectrum is inherited analytically, while the spanwise (cylindrical-cross-section) dependence is expanded by a Weyl--Sommerfeld identity for the Hankel harmonics $H_m^{(1)}(kr)\ee^{\ii m\theta}$, valid in the half-space containing the plate. The second stage is therefore exact. Superposition over the spectrum yields the far-field directivity of the scattered field. We note that this is a one-way coupling: we assume the scattered field is not re-scattered by the duct, an approximation consistent with the plate being relatively far from the duct exit. Should this one-way approximation prove insufficient in other configurations, a straightforward iterative scheme between the two stages can be employed to recover the full coupling.

On the other hand, the  LLM-assisted theoretical research has produced visible successes in the last two years, from mathematical discovery by program search \citep{romera2024} to olympiad-level geometry \citep{trinh2024} and interactive derivation assistance in applied mathematics \citep{liu2024}. We state plainly, however, that the core contribution of the paper is not the use of an LLM but our recognition that a Weyl--Sommerfeld identity translates exactly between the cylindrical modal waves of the duct and the Cartesian gust incident on the plate, so that two classical Wiener--Hopf solutions can be chained without approximation. That idea, and the chaining strategy, are from human. Because both modelling stages were already available as validated implementations, the most important (and theoretically difficult) kernel factorisations, the pole tracking and the causality arguments are taken from our previous works \citep{huang2016,huang2017} -- which sidesteps the present immaturity of LLMs in rigorous complex-variable reasoning. The LLM proved particularly helpful for the tedious algebraic manipulations required to bridge the two coordinate systems and for the subsequent verifications detailed in §\ref{sec:numerics}. 
We expect this division of labour -- human supplies the key mathematical idea, the LLM the mechanical execution under systematic verification -- to be a productive pattern for similar multi-model coupling problems.

The remaining part of this paper is organised as follows. Section~\ref{sec:model} states the geometry and conventions and develops the model: the stage-one duct solution is summarised in \S\ref{sec:stage1}, the spectral bridge by LLM in \S\ref{sec:bridge}, the stage-two plate Wiener--Hopf problem in \S\ref{sec:stage2}, and the far-field asymptotics, including the image-source identity, in \S\ref{sec:farfield}. Readers who are interest mainly in the underlying wave mechanics may directly skip this section and proceed to the  results and discussions in the following sections. Section~\ref{sec:numerics} describes the numerical implementation and the validation details. Section~\ref{sec:results} presents near- and far-field results for a spinning mode $(m,n)=(5,1)$ at normalised angular frequency $\omega=20$, in both a fully quiescent and a with-jet configuration, and the full-circle directivity of a higher-order mode $(13,1)$ at $\omega=28.3$. Section~\ref{sec:conclusions} concludes the whole work.

\section{Theoretical model}
\label{sec:model}

\subsection{Statement of the problem}
\label{sec:geometry}

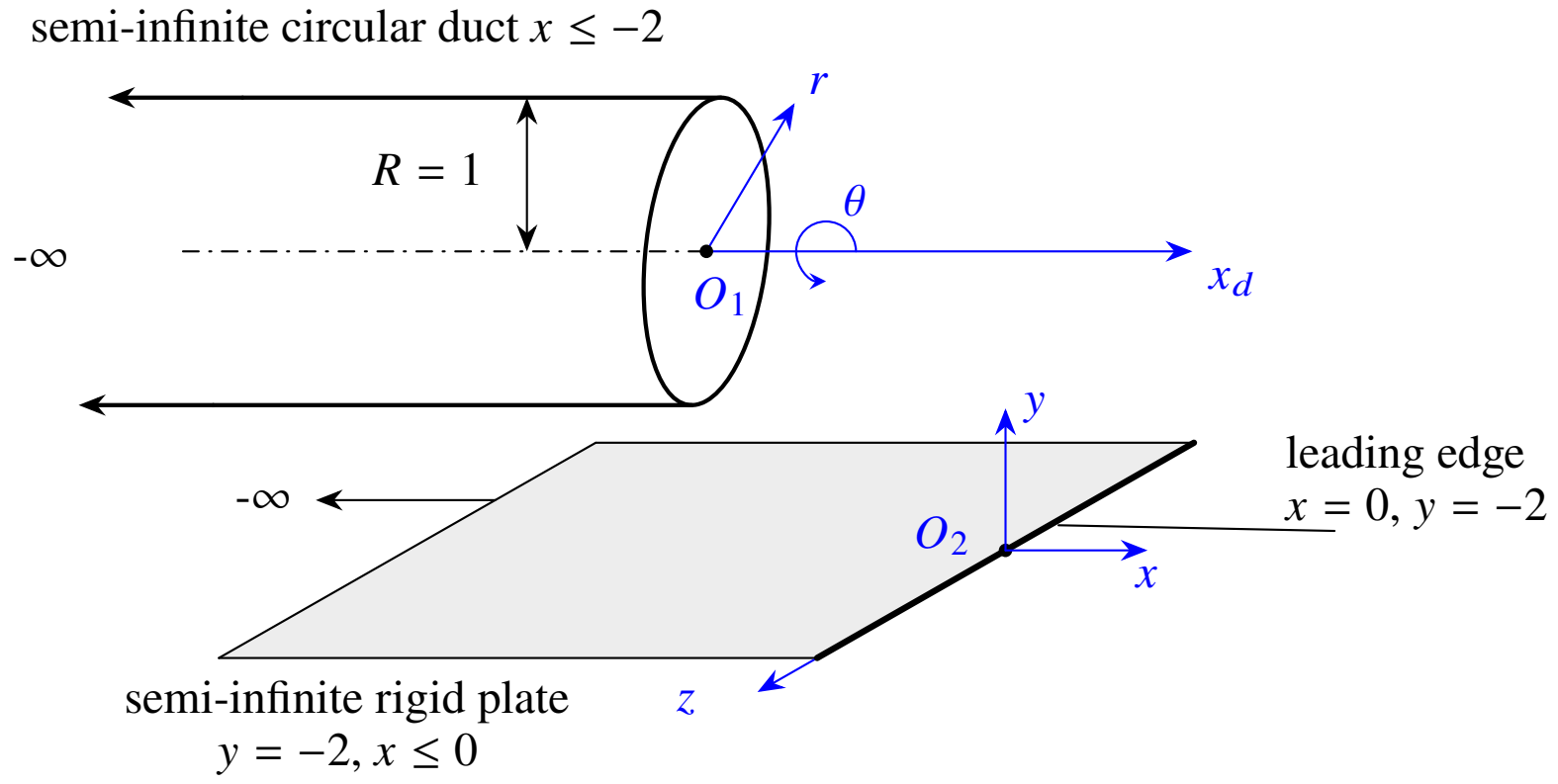
\begin{figure}
  \centering
\begin{tikzpicture}[x={(1cm,0cm)}, y={(0cm,1cm)}, z={(0.42cm,0.24cm)},
    line join=round, line cap=round, >={Stealth[length=1.8mm,width=1.4mm]}]

  \fill[gray!14] (-4,-2,-3) -- (-4,-2,3) -- (0,-2,3) -- (0,-2,-3) -- cycle;
  \draw[thin] (-4,-2,-3) -- (0,-2,-3);
  \draw[thin] (-4,-2,3) -- (0,-2,3);
  \draw[thin] (-4,-2,-3) -- (-4,-2,3);
  \draw[thin,->] (-4,-2,1.4) -- (-5.2,-2,1.4);				
  \node at (-5.55,-2,1.4) {-$\infty$};
  \draw[blue,thin,->] (0,-2,-3) -- (0,-2,-3.95);			
  \node at (-0.35,-2,-4.25) {\textcolor{blue}{$z$}};		
  \draw[very thick] (0,-2,-3) -- (0,-2,3);					
  \node[align=left] at (2.75,-1.55,0) {leading edge\\ $x=0$, $y=-2$};
  \draw[thin] (2.2,-1.87,0) -- (0.06,-2,0.7);				
  \node[align=center] at (-3.9,-2.9,-1.2) {semi-infinite rigid plate\\ $y=-2$, $x\le0$};

  \draw[thick, canvas is yz plane at x=-2] (0,0) circle (1);
  \draw[thick] (-2,0.9724,0.2334) -- (-5.2,0.9724,0.2334);
  \draw[thick] (-2,-0.9724,-0.2334) -- (-5.2,-0.9724,-0.2334);
  \draw[thick,->] (-5.2,0.9724,0.2334) -- (-6.1,0.9724,0.2334);
  \draw[thick,->] (-5.2,-0.9724,-0.2334) -- (-6.1,-0.9724,-0.2334);
  \node at (-6.45,-0.05,0) {-$\infty$};
  \draw[dash dot, thin] (-5.5,0,0) -- (-2.04,0,0);
  \node[align=center] at (-4.4,1.5,0) {semi-infinite circular duct  $x\le-2$};
  \draw[blue,->] (-2,0,0) -- (1.25,0,0) node[below right] {$x_d$};
  \draw[blue,->] (-2,0,0) -- (-1.4098,0.9922,0) ;
  \node at (-1.24,1.12,0) {\textcolor{blue}{$r$}};
      	\draw[blue,-stealth] (-1,0) arc (0:270:0.2);	
	\draw[blue](-1.2,0.34,0) node [right]{$\theta$};	
	
	\draw[blue](-2.2,-0.3,0) node [right]{$O_1$};
	  \fill (-2,-0,0) circle (0.045);	
	
  \fill (0,-2,0) circle (0.045);
  \node[left] at (-0.12,-1.9,0) {\textcolor{blue}{$O_2$}};
  \draw[blue,->] (0,-2,0) -- (0.95,-2,0) node[below] {$x$};
  \draw[blue,->] (0,-2,0) -- (0,-1.05,0) node[right] {$y$};	

  \draw[thin,<->] (-3.2,0,0) -- (-3.298,0.9724,0.2334);		
  \node[left] at (-3.38,0.55,0) {$R=1$};

\end{tikzpicture}
  \caption{Geometry of the coupled problem. The semi-infinite circular duct of radius $R=1$ occupies $x\le-2$; the duct-local cylindrical coordinates $(x_d,r,\theta)$ have their origin at the exit centre $O_1=(-2,0,0)$, with the $x_d$ axis along the duct axis and the $r$, $\theta$ directions as drawn on the exit face. The semi-infinite rigid plate of vanishing thickness lies in the plane $y=-2$ and occupies $x\le0$; its straight leading edge $x=0$ is marked by the thick line. The plate-local Cartesian frame is attached at $O_2=(0,-2,0)$. The streamwise gap between the exit plane and the edge is $L=2$ and the vertical offset is $d=2$. }
  \label{fig:setup}
\end{figure}
In this problem set-up, all lengths are non-dimensionalised by the duct radius $D$, so that the duct radius is unity and the dimensionless frequency is $\omega=2\piup fD/c_0$ with $f$ the physical frequency and $c_0$ the speed of sound. The time dependence $\mathrm{exp}({-\ii\omega t})$ is suppressed throughout for brevity. A global Cartesian frame $(x,y,z)$ is used, with $x$ along the duct axis and the plate (positive downstream), $y$ normal to the plate (positive towards the duct side) and $z$ spanwise. The duct is semi-infinite, with wall $r=1$ and occupying $x\le-2$; the exit plane is $x=-2$, with exit centre $(x,r)=(-2,0)$. The duct may be annular with a centrebody of radius $r=h$, the hollow duct corresponding to $h=0$ (which is assumed in the rest of this paper). Cylindrical coordinates $(r,\theta)$ are taken about the duct axis, $r=\sqrt{y^2+z^2}$, $\theta=\operatorname{atan2}(z,y)$ measured from the $+y$ axis. The duct-local axial coordinate is $x_d=x+2$ (the subscript $d$ stands for `duct' and marks duct-frame quantities throughout), so that the exit is $x_d=0$.

The plate is rigid, of vanishing thickness and semi-infinite: it lies in the plane $y=-d$ with $d=2$ for the rest studies, occupies $x\le0$, and its straight leading edge is the line $x=0$, $y=-d$ (the $z$ axis shifted by $-d$). We write $\tilde y=y+d$, so that the plate surface is $\tilde y=0$; the distance between the exit plane and the edge is $L=2$. Consequently, the plate lies outside the duct and outside the shed vortex sheet, in the uniform ambient medium. The geometry, the two coordinate systems and the whole set-up are summarised in figure~\ref{fig:setup}.

The configuration consists of a single cylindrical duct (no centrebody, $h=0$) discharging a subsonic jet of Mach number $M_1$ into a quiescent ambient ($M_0=0$), for simplicity the jet-to-ambient density ratio being unity; the fully quiescent limit $M_1=0$ is contained as a special case. The stage-one solution includes the jet and the vortex sheet shed from the duct lip, with a Kutta condition applied at the lip; the jet enters the coupling only through the output spectrum $F_+(u)$. The stage-two plate problem, by contrast, treats the medium surrounding the plate as quiescent and uniform: the downstream vortex sheet is presumably not advected over the plate. This is a deliberate one-way-coupling simplification, discussed further in \S\ref{sec:conclusions}. A uniform external flow $M_0\neq0$ is accommodated stage by stage by the Lorentz transformation of \citet{gabard2006} and the Prandtl--Glauert transformation of \citet{huang2017}, and is not pursued here. Three configurations are computed in \S\ref{sec:results}: a fully quiescent baseline ($M_0=M_1=0$), in which no vortex sheet is shed and no shear-layer dynamics is present -- chosen deliberately to remove the shear layer from the coupled problem, so that the model can later be validated directly against numerical simulation of the pure acoustic radiation--scattering process -- and two with-jet cases ($M_1= 0.353$ and $ 0.338$). A plug-flow velocity profile is adopted for the jet and the ambient stream, which is a common idealisation in aeroacoustic modelling of aeroengine noise radiation (e.g., \citet{munt1977, rienstra1984}). The present configurations are idealised from our prior industrial problems for the Trent-series aeroengine \citep{huang2008}.

Two different Fourier conventions are used, one per stage, matching the source formulations from their original references:
\begin{enumerate}
  \item stage 1 (duct, the convention of \citet{gabard2006}): forward kernel $\mathrm{exp}({-\ii\omega u x_d})$, inverse transform $\psi=(\omega/2\piup)\int_\Gamma\beta\,\ee^{+\ii\omega u x_d}\dd u$;
  \item stage 2 (plate, the convention of \citet{huang2017}): forward kernel $\mathrm{exp}({+\ii\alpha x})$, inverse transform $\psi=(1/2\piup)\int\beta\,\ee^{-\ii\alpha x}\dd\alpha$.
\end{enumerate}
Consequently, an axial dependence $\mathrm{exp}({+\ii\omega u x})$ produced by the duct solution corresponds, in the plate problem, to the incident pole $\alpha_p=-\omega u$. This sign bookkeeping is essential and is verified numerically in \S\ref{sec:numerics}. The other main symbols are collected in table~\ref{tab:symbols}.

\begin{table}
\centering
\begin{tabular}{ll}
Symbol & Meaning \\
\midrule
$(m,\mu)$ & incident duct mode: azimuthal order $m$, axial wavenumber $\omega\mu$ \\
$x_d$ & duct-local axial coordinate, the exit is $x_d=0$ \\
$\alpha_{mn}$ & internal radial wavenumber, root of the eigencondition $\Lambda(\alpha)=0$ \\
$F_+(u)$ & stage-one Wiener--Hopf solution (half-range transform of the vortex-sheet displacement) \\
$K(u),\tilde K_\pm(u)$ & stage-one Wiener--Hopf kernel and its factorisation \\
$\lambda_0(u)$ & external radial wavenumber over $\omega$; $\lambda_0=\sqrt{1-u^2}$ when $M_0=0$ \\
$u,\nu,\alpha$ & spectral variables: axial (duct convention), spanwise, streamwise (plate convention) \\
$\gamma_d(\nu)$ & $\sqrt{(\lambda_0\omega)^2-\nu^2}$, wall-normal wavenumber of an incident component \\
$\khat,\gamma_p(\alpha)$ & $\khat=\sqrt{\omega^2-\nu^2}$; $\gamma_p=\sqrt{\alpha^2-\khat^2}$, stage-two kernel \\
$\gamma_\pm(\alpha)$ & $\sqrt{\alpha\pm\khat}$, explicit factorisation of the stage-two kernel \\
$M_0$ & the Mach number of the ambient environment \\
$M_1$ & the Mach number of the jet flow from the duct \\
\end{tabular}
\caption{Principal symbols.}
\label{tab:symbols}
\end{table}

\subsection{Stage 1: radiation from the duct exit}
\label{sec:stage1}

The first stage is the duct-radiation model of \citet{munt1977}, \citet{rienstra1984} and \citet{gabard2006}, used here without modification; we nevertheless rederive it in some detail, because the notation and the output quantity $F_+(u)$ are the interface on which the whole coupled model rests. The derivation follows \citet{gabard2006}, equations of which are cited as (G3.$n$).

\subsubsection{Governing equations and boundary, edge and radiation conditions}
\label{sec:stage1eqs}

The incident duct mode, normalised as in appendix~B of \citet{gabard2006}, is
\begin{equation}
\phi_0=\Psi(r)\,\ee^{\ii\omega\mu x_d+\ii m\theta},\qquad
\Psi(r)=Y_m'(\alpha h)J_m(\alpha r)-J_m'(\alpha h)Y_m(\alpha r),
\label{eq:incmode}
\end{equation}
\noindent where the radial wavenumber $\alpha=\alpha_{mn}$ satisfies the annular-duct eigencondition
\begin{equation}
\Lambda(\alpha)\equiv Y_m'(\alpha h)J_m'(\alpha)-J_m'(\alpha h)Y_m'(\alpha)=0,
\label{eq:eigen}
\end{equation}
which reduces to $J_m'(\alpha)=0$ for the hollow duct ($h=0$); the radial mode number $n=1,2,\dots$ counts the roots in increasing order, $n=1$ being the lowest radial solution. Since the geometry and the boundary conditions are axisymmetric, the diffracted field retains the azimuthal and time dependence of the incident mode; writing the total field as
\begin{equation}
\phi=\big[\psi(r,x_d)+\psi_0(r,x_d)\big]\,\ee^{\ii m\theta-\ii\omega t},
\qquad
\psi_0(r,x_d)=\Psi(r)\,\ee^{\ii\omega\mu x_d},
\label{eq:decomp}
\end{equation}
\noindent and the radial displacement of the shed vortex sheet as $\xi(x_d)\,\mathrm{exp}({\ii m\theta-\ii\omega t})$, the diffracted potential $\psi$ solves the convected Helmholtz equations (G2.11)--(G2.12)
\begin{align}
\frac{1}{r}\partial_r\big(r\partial_r\psi\big)+\partial_{x_d}^2\psi-\frac{m^2}{r^2}\psi
-\big(-\ii\omega+M_0\partial_{x_d}\big)^2\psi&=0,
&& r>1,
\label{eq:convect0}\\
\frac{1}{r}\partial_r\big(r\partial_r\psi\big)+\partial_{x_d}^2\psi-\frac{m^2}{r^2}\psi
-C_1^2\big(-\ii\omega+M_1\partial_{x_d}\big)^2\psi&=0,
&& h<r<1,
\label{eq:convect1}
\end{align}
\noindent where $C_1=c_1/c_0$ is the jet-to-ambient sound-speed ratio ($C_1$ is simply set to 1 in all our cases) and $(-\ii\omega+M\partial_{x_d})$ is the material derivative for the $\mathrm{exp}({-\ii\omega t})$ convention. The boundary conditions are (G2.13)--(G2.17): the hard-wall condition on the duct and on the centrebody,
\begin{equation}
\partial_r\psi(1^-,x_d)=\partial_r\psi(1^+,x_d)=0\quad(x_d\le0);
\qquad
\partial_r\psi(h,x_d)=0;
\label{eq:hardwall}
\end{equation}
the kinematic conditions coupling each side of the vortex sheet to its displacement for $x_d>0$,
\begin{equation}
\big(-\ii\omega+M_0\partial_{x_d}\big)\xi=\partial_r\psi(1^+,x_d),
\qquad
\big(-\ii\omega+M_1\partial_{x_d}\big)\xi=\partial_r\psi(1^-,x_d);
\label{eq:kinematic}
\end{equation}
and the continuity of pressure across the vortex sheet,
\begin{equation}
\big(-\ii\omega+M_0\partial_{x_d}\big)\psi(1^+,x_d)
=D_1\big(-\ii\omega+M_1\partial_{x_d}\big)\big[\psi(1^-,x_d)+\psi_0(1^-,x_d)\big],
\quad x_d>0,
\label{eq:pcont}
\end{equation}
where $D_1=\rho_1/\rho_0$ is the jet-to-ambient density ratio ($D_1=1$ in the baseline case) and the pressure is $p=-\rho(-\ii\omega+M\partial_{x_d})\psi$ in each medium, so that in the quiescent ambient $p_d=\ii\omega\psi_d$ (an overall sign convention is immaterial here). Note that the incident wave $\psi_0$ appears in (\ref{eq:pcont}) only: the incident mode is therefore equivalent to a prescribed pressure-source distribution on the continuation $r=1$, $x_d>0$ of the duct wall, which is whatwhat makes the Wiener--Hopf formulation below possible.

Two further classes of conditions close the problem. First, the {edge conditions} at the duct lip $(r,x_d)=(1,0)$: integrability of the pressure requires the potential to be finite at the lip, but the amount of vorticity shed into the sheet must be prescribed in addition. If no vorticity is shed, $\psi(1,x_d)=O(x_d^{1/2})$ as $x_d\to0^+$ and the pressure is singular; the full Kutta condition (all vorticity shed) requires $\psi(1,x_d)=O(x_d^{3/2})$, i.e.\ finite pressure at the lip. Intermediate situations are parametrised by Rienstra's shedding parameter $\gamma_K\in[0,1]$ introduced in (\ref{eq:Fp}) below. Secondly, the {radiation condition} is enforced through causality: the analysis is carried out with a complex frequency $\omega=|\omega|\ee^{\ii\epsilon}$, $0\le\epsilon\le\piup/2$, for which all waves, including the Kelvin--Helmholtz instability wave of the vortex sheet, decay at infinity; the real-frequency solution is then recovered as the limit $\epsilon\to0$ with analytic continuation, which amounts to indenting the inversion contour around the instability zero (see below).

\subsubsection{Fourier transform and the Wiener--Hopf equation}
\label{sec:stage1Wiener--Hopf}

The axial Fourier transform and its half-range parts are defined by
\begin{equation}
\begin{aligned}
\beta(r,u)&=\int_{-\infty}^{+\infty}\psi(r,x_d)\,\ee^{-\ii\omega u x_d}\dd x_d,
\qquad
\beta=\beta_-+\beta_+,\\
\beta_-(r,u)&=\int_{-\infty}^{0}\!\!\cdots\;,
\qquad
\beta_+(r,u)=\int_{0}^{+\infty}\!\!\cdots\;,
\end{aligned}
\label{eq:ft1}
\end{equation}
with $\omega u$ the axial wavenumber. With $\epsilon$ close to $\piup/2$, $\beta_-$ and $\beta_+$ are regular in overlapping half-planes whose intersection is a strip $\mathcal S$ containing the real axis; the physical solution is recovered by deforming the contour within $\mathcal S$ as $\epsilon\to0$. Applying the transform to (\ref{eq:convect0})--(\ref{eq:convect1}) gives the Bessel equations (G3.2)--(G3.3)
\begin{equation}
\frac{1}{r}\partial_r\big(r\partial_r\beta\big)
+\Big(\omega^2\lambda_{0,1}^2-\frac{m^2}{r^2}\Big)\beta=0,
\qquad
\lambda_0^2=(1-uM_0)^2-u^2,\quad
\lambda_1^2=C_1^2(1-uM_1)^2-u^2,
\label{eq:bessel}
\end{equation}
in the ambient and in the jet respectively, where $\lambda_{0,1}$ are the radial wavenumbers divided by $\omega$. They are written in product form $\lambda_{0}=\lambda_0^-\lambda_0^+$, $\lambda_1=\lambda_1^-\lambda_1^+$ with
\begin{equation}
\lambda_0^\pm=\sqrt{1-u(M_0\pm1)},
\qquad
\lambda_1^\pm=\sqrt{C_1-u(M_1C_1\pm1)},
\label{eq:lambdapm}
\end{equation}
principal branches being taken, so that $\lambda_0^\pm$ (resp.\ $\lambda_1^\pm$) are regular and non-zero in the half-planes $R_0^\pm$ (resp.\ $R_1^\pm$) whose intersection is $\mathcal S$; in $\mathcal S$ one has $\mathrm{Im}\,\lambda_{0,1}\omega>0$. The solution of (\ref{eq:bessel}) that satisfies the centrebody condition in (\ref{eq:hardwall}) and decays as $r\to+\infty$ is (G3.5)
\begin{equation}
\beta(r,u)=
\begin{cases}
A(u)\,H_m^{(1)}(\lambda_0\omega r), & r>1,\\[2pt]
B(u)\big[Y_m'(\lambda_1\omega h)J_m(\lambda_1\omega r)
-J_m'(\lambda_1\omega h)Y_m(\lambda_1\omega r)\big], & h<r<1.
\end{cases}
\label{eq:betasol1}
\end{equation}
Let $F_+(u)=\int_0^\infty \xi(x_d)\,\ee^{-\ii\omega u x_d}\dd x_d$ be the half-range transform of the vortex-sheet displacement. The transformed kinematic conditions (\ref{eq:kinematic}) then give (G3.6)--(G3.7)
\begin{align}
-\ii(1-uM_0)F_+(u)&=A(u)\,\lambda_0\,H_m^{(1)\prime}(\lambda_0\omega),
\label{eq:relA}\\
-\ii(1-uM_1)F_+(u)&=B(u)\,\lambda_1\big[Y_m'(\lambda_1\omega h)J_m'(\lambda_1\omega)
-J_m'(\lambda_1\omega h)Y_m'(\lambda_1\omega)\big],
\label{eq:relB}
\end{align}
where $\partial_{x_d}\to\ii\omega u$ and the factor $\omega$ of the transform convention has been divided out. Next, introduce the transform of the pressure mismatch across the outer duct wall ($x_d<0$) and the vortex sheet ($x_d>0$),
\begin{align}
G(u)&=\int_{-\infty}^{+\infty}
\Big[\big(-\ii\omega+M_0\partial_{x_d}\big)\psi(1^+,x_d)
-D_1\big(-\ii\omega+M_1\partial_{x_d}\big)\psi(1^-,x_d)\Big]
\ee^{-\ii\omega u x_d}\dd x_d
\nonumber\\
&=G_-(u)+G_+(u).
\label{eq:Gdef}
\end{align}
On the one hand, substituting (\ref{eq:betasol1}) into (\ref{eq:Gdef}) gives (G3.8)
\begin{align}
G(u)&=-\ii\omega\Big\{
(1-uM_0)A(u)H_m^{(1)}(\lambda_0\omega)
\nonumber\\
&\qquad
-D_1(1-uM_1)B(u)\big[Y_m'(\lambda_1\omega h)J_m(\lambda_1\omega)
-J_m'(\lambda_1\omega h)Y_m(\lambda_1\omega)\big]\Big\}.
\label{eq:GAB}
\end{align}
On the other hand, the pressure continuity (\ref{eq:pcont}) determines $G_+$ explicitly in terms of the incident wave:
\begin{align}
G_+(u)
&=\int_0^\infty D_1\big(-\ii\omega+M_1\partial_{x_d}\big)\psi_0(1^-,x_d)\,\ee^{-\ii\omega u x_d}\dd x_d
\nonumber\\
&=-\ii\omega D_1\Psi(1)(1-\mu M_1)\int_0^\infty \ee^{\ii\omega(\mu-u)x_d}\dd x_d
=-\ii\omega D_1\Psi(1)(1-\mu M_1)\,\frac{\ii}{\omega(\mu-u)},
\nonumber\\
&=\frac{D_1\Psi(1)(1-\mu M_1)}{\mu-u},
\label{eq:Gplus}
\end{align}
where $\int_0^\infty \ee^{\ii\omega(\mu-u)x_d}\dd x_d=\ii/[\omega(\mu-u)]$ holds for $u$ in the strip $\mathcal S$ (the causal $\mathrm{Im}\,\omega>0$ provides convergence at $x_d\to+\infty$), and $(-\ii)(\ii)=1$. Eliminating $A$ and $B$ between (\ref{eq:relA}), (\ref{eq:relB}) and (\ref{eq:GAB}) yields the Wiener--Hopf equation (G3.10)
\begin{equation}
G_-(u)+G_+(u)=\frac{\omega K(u)}{\lambda_0\lambda_1}\,F_+(u),
\label{eq:Wiener--Hopf1}
\end{equation}
with the kernel (G3.11)
\begin{align}
K(u)&=
D_1(1-uM_1)^2\lambda_0\,
\frac{Y_m'(\lambda_1\omega h)J_m(\lambda_1\omega)-J_m'(\lambda_1\omega h)Y_m(\lambda_1\omega)}
{Y_m'(\lambda_1\omega h)J_m'(\lambda_1\omega)-J_m'(\lambda_1\omega h)Y_m'(\lambda_1\omega)}
\nonumber\\
&\quad
-(1-uM_0)^2\lambda_1\,
\frac{H_m^{(1)}(\lambda_0\omega)}{H_m^{(1)\prime}(\lambda_0\omega)} .
\label{eq:kernel}
\end{align}
In (\ref{eq:Wiener--Hopf1}), $G_+$, $K$, $\lambda_0$ and $\lambda_1$ are known, while $G_-$ (regular in $R^-$) and $F_+$ (regular in $R^+$) are to be found. The kernel plays the role of the dispersion relation of the problem: its poles, located at the zeros of the denominator of the first term of (\ref{eq:kernel}), are precisely the eigencondition (\ref{eq:eigen}) of the infinite annular duct, and its zeros include the double zero $u_0$ of vorticity shedding for a uniform flow (a simple zero describing the instability wave for a non-uniform flow).

\subsubsection{Factorisation of the kernel and solution for $F_+$}
\label{sec:stage1sol}

The kernel is factored as (G3.12)
\begin{equation}
K(u)=\frac{\tilde K_+(u)}{\tilde K_-(u)}\,(u-u_0)^b,
\label{eq:Ksplit}
\end{equation}
where $\tilde K_\pm$ are regular and non-zero in $R^\pm$; the zero at $u_0$ is factored explicitly, with $b=2$ for a uniform flow (double zero, vortex shedding) and $b=1$ otherwise (simple zero, instability wave). No closed-form split exists here, and $\tilde K_\pm$ are computed numerically from the general Cauchy formula (G3.23)
\begin{equation}
\log \tilde K_\pm(u)=
-\frac{1}{2\ii\piup}\int_{C_\pm}
\frac{\log\big[K(\zeta)/(\zeta-u_0)^b\big]}{\zeta-u}\,\dd\zeta,
\label{eq:cauchy}
\end{equation}
with $C_\pm$ running through the strip $\mathcal S$, $C_+$ passing above and $C_-$ below the point $u$. Multiplying (\ref{eq:Wiener--Hopf1}) by $\lambda_0^-\lambda_1^-\tilde K_-$ and using $\lambda_{0,1}=\lambda_{0,1}^-\lambda_{0,1}^+$ together with (\ref{eq:Ksplit}) gives
\begin{equation}
\frac{\omega F_+(u)\tilde K_+(u)(u-u_0)^b}{\lambda_0^+(u)\lambda_1^+(u)}
=\lambda_0^-\lambda_1^-\tilde K_-(u)\,G_-(u)
+\lambda_0^-\lambda_1^-\tilde K_-(u)\,G_+(u).
\label{eq:Wiener--Hopf1b}
\end{equation}
The left-hand side is regular in $R^+$ and the first term on the right in $R^-$, but the last term is not regular in $R^-$ because $G_+$ has the simple pole at $u=\mu$; subtracting its value at $u=\mu$ on both sides repairs this (G3.13):
\begin{align}
\frac{\omega F_+(u)\tilde K_+(u)(u-u_0)^b}{\lambda_0^+(u)\lambda_1^+(u)}
+\lambda_0^-(\mu)\lambda_1^-(\mu)\tilde K_-(\mu)G_+(u)
&=\lambda_0^-\lambda_1^-\tilde K_-(u)G_-(u)
\nonumber\\
&\quad
+G_+(u)\big[\lambda_0^-(u)\lambda_1^-(u)\tilde K_-(u)
\nonumber\\
&\qquad
-\lambda_0^-(\mu)\lambda_1^-(\mu)\tilde K_-(\mu)\big].
\label{eq:Wiener--Hopf1c}
\end{align}
Both sides are now regular in $R^+$ and $R^-$ respectively, hence together define an entire function $E(u)$. For the outlet problem with the full Kutta condition, the edge condition $\psi(1,x_d)=O(x_d^{3/2})$ translates into $F_+(u)=O(u^{-5/2})$ as $|u|\to\infty$ in $\mathcal S$; with $G_+=O(u^{-1})$, $\lambda_{0,1}^\pm=O(u^{1/2})$ and $\tilde K_\pm=O(u^{\pm1/2})$ one finds $E(u)=O(u^{-1})$, so $E\equiv0$ by Liouville's theorem. Setting the left-hand side of (\ref{eq:Wiener--Hopf1c}) to zero then gives
\begin{equation}
F_+(u)=
-\frac{\lambda_0^-(\mu)\lambda_1^-(\mu)\tilde K_-(\mu)\,\lambda_0^+(u)\lambda_1^+(u)}
{\omega(u-u_0)^b\,\tilde K_+(u)}\,G_+(u).
\label{eq:Fpkutta}
\end{equation}
Substituting $G_+$ from (\ref{eq:Gplus}) and writing the result in Rienstra's form with the shedding parameter $\gamma_K\in[0,1]$ (which interpolates between no shedding, $\gamma_K=0$, and the full Kutta condition, $\gamma_K=1$), the stage-one solution is (G3.19)
\begin{equation}
F_+(u)=\frac{\Psi(1)D_1(1-\mu M_1)\,\lambda_0^-(\mu)\lambda_1^-(\mu)\tilde K_-(\mu)\,
\lambda_0^+(u)\lambda_1^+(u)}
{\omega\,(u-u_0)^b(\mu-u_0)\,\tilde K_+(u)}
\left(\frac{u-u_0}{u-\mu}-\gamma_K\right);
\label{eq:Fp}
\end{equation}
indeed, for $\gamma_K=1$ the bracket reduces to $(\mu-u_0)/(u-\mu)$ and (\ref{eq:Fp}) recovers (\ref{eq:Fpkutta}) with (\ref{eq:Gplus}).

\subsubsection{The radiated field and the duct-alone directivity}
\label{sec:stage1field}

The diffracted field follows from the inverse transform of (\ref{eq:betasol1}) with $A$ eliminated by (\ref{eq:relA}):
\begin{align}
\psi_d(r,\theta,x_d)&=\ee^{\ii m\theta}\,\frac{\omega}{2\ii\piup}\int_\Gamma
\frac{(1-uM_0)F_+(u)}{\lambda_0(u)}\,
T_0(r,u)\,\ee^{\ii\omega u x_d}\,\dd u,
\qquad r>1,
\nonumber\\
T_0(r,u)&=\frac{H_m^{(1)}(\lambda_0\omega r)}{H_m^{(1)\prime}(\lambda_0\omega)},
\label{eq:ductfieldfull}
\end{align}
and $p_d=\ii\omega\psi_d$ in the quiescent ambient. The contour $\Gamma$ lies in the strip $\mathcal S$; as $\epsilon\to0$ it runs from $-\infty+0\ii$ to $+\infty-0\ii$, crosses the real axis at $u_c=M_1C_1^2/(M_1^2C_1^2-1)$, and is indented to pass below the instability zero $u_0$, in accordance with causality: the zero crosses the contour as $\epsilon$ decreases, and the indentation is what remains of the causal prescription. The residue of the pole at $u_0$ in (\ref{eq:ductfieldfull}) is the Kelvin--Helmholtz instability wave; in the present baseline model this contribution, which decays radially, is not fed into the second stage (see the assumptions discussed in \S\ref{sec:conclusions}).

For the quiescent ambient of the baseline case ($M_0=0$, $\lambda_0=\sqrt{1-u^2}$), (\ref{eq:ductfieldfull}) reduces to
\begin{equation}
\psi_d(r,\theta,x_d)=\ee^{\ii m\theta}\,\frac{\omega}{2\piup}\int_\Gamma
\frac{-\ii F_+(u)}{\lambda_0(u)}\,
\frac{H_m^{(1)}(\lambda_0\omega r)}{H_m^{(1)\prime}(\lambda_0\omega)}\,
\ee^{\ii\omega u x_d}\,\dd u,
\qquad \lambda_0(u)=\sqrt{1-u^2}.
\label{eq:ductfield}
\end{equation}
The far field of the duct alone, needed in \S\ref{sec:image}, follows from (\ref{eq:ductfield}) with $p_d=\ii\omega\psi_d$ by standard asymptotics. With $r=R\sin\Theta$, $x_d=R\cos\Theta$ ($\Theta$ measured from the duct axis), the Hankel function is replaced by its large-argument form $H_m^{(1)}(w)\sim\sqrt{2/(\piup w)}\,\ee^{\ii(w-m\piup/2-\piup/4)}$, and the phase
\begin{equation}
\Phi(u)=\omega\big(\lambda_0(u)\,r+u x_d\big),\qquad
\Phi'(u)=\omega\Big(-\frac{u}{\lambda_0}r+x_d\Big),\qquad
\Phi''(u)=-\frac{\omega r}{\lambda_0^3},
\end{equation}
has its saddle at $u_s=\cos\Theta$, where $\lambda_0(u_s)=\sin\Theta$ and $\Phi(u_s)=\omega R$. Since $\Phi''(u_s)=-\omega R/\sin^2\Theta<0$, the stationary-phase formula $\int g\,\ee^{\ii\Phi}\dd u\sim g(u_s)\sqrt{2\piup/|\Phi''(u_s)|}\,\ee^{\ii\Phi(u_s)-\ii\piup/4}$ gives
\begin{equation}
p_d\sim D_P(\Theta)\,\ee^{\ii m\theta}\,\frac{\ee^{\ii\omega R}}{R},
\qquad
D_P(\Theta)=
-\frac{\ii\omega}{\piup}\,\ee^{-\ii m\piup/2}\,
\frac{F_+(\cos\Theta)}{\sin\Theta\;H_m^{(1)\prime}(\omega\sin\Theta)},
\label{eq:ductdir}
\end{equation}
where the two square-root factors simplify as
$\sqrt{2/(\piup\omega R\sin^2\Theta)}\times\sqrt{2\piup\sin^2\Theta/(\omega R)}=2/(\omega R)$
and $\ee^{-\ii\piup/4}\cdot\ee^{-\ii\piup/4}=-\ii$. Equation (\ref{eq:ductdir}) is the directivity computed by the stage-one code (\texttt{farfield.m}), and is the quantity $D_P$ appearing in the image identity (\ref{eq:image}) below.

The interface between the two stages of the present model is precisely the function $F_+(u)$: nothing in the stage-one derivation or code is modified.

\subsection{The bridge from cylindrical to Cartesian coordinates}
\label{sec:bridge}

The first stage delivers the incident field in the form (\ref{eq:ductfield}): a one-dimensional axial spectrum, known on the contour $\Gamma$, of cylindrical harmonics $H_m^{(1)}(\lambda_0\omega r)\mathrm{exp}({\ii m\theta})$. The plate problem of \S\ref{sec:stage2}, by contrast, is closed only as a two-dimensional plane-wave spectrum $\widehat A(\alpha,\nu)$ of convected gusts $\mathrm{exp}({\ii\alpha x+\ii\nu z})$. The present subsection constructs the exact conversion between these two representations -- the mathematical heart of the coupling. 

Two structural facts make the conversion possible without any approximation. First, the duct and the plate share translation invariance in $x$: the axial Fourier variable is common to both problems, and on the plate plane the axial spectrum is simply inherited from the duct solution, $\alpha=\omega u$ (\S\ref{sec:axial}). Secondly, the remaining cylindrical dependence is converted by a Weyl--Sommerfeld angular-spectrum identity (\S\ref{sec:weyl}), in which the whole azimuthal structure of the spinning mode is carried by the algebraic factor $((\gamma-\ii\nu)/k)^m$: the identity is exact, holds component by component along the contour $\Gamma$, and turns the modal radiation of the duct into the incoming gust of the plate. The output is the double spectrum of the incident normal velocity on the plate plane (\S\ref{sec:incspectrum}), which is the exact input to the second Wiener--Hopf stage for the plate scattering.

\subsubsection{The axial spectrum is inherited}
\label{sec:axial}

On the plate plane $y=-d$ a field point has
\begin{equation}
r_p(z)=\sqrt{d^2+z^2},\qquad \theta_p(z)=\operatorname{atan2}(z,-d),\qquad x_d=x+L,
\end{equation}
so that (\ref{eq:ductfield}) gives
\begin{equation}
\psi_d\big|_{\tilde y=0}=\frac{\omega}{2\piup}\int_\Gamma
\frac{-\ii F_+(u)}{\lambda_0(u)}\,
\frac{H_m^{(1)}(\lambda_0\omega r_p)}{H_m^{(1)\prime}(\lambda_0\omega)}\,
\ee^{\ii m\theta_p(z)}\,\ee^{\ii\omega uL}\,\ee^{\ii\omega ux}\,\dd u .
\label{eq:onplate}
\end{equation}
Each $u$ component is therefore already a streamwise gust $\ee^{\ii\omega ux}$ on the plate: the axial spectrum is known analytically and requires no numerical transform.

\subsubsection{A Weyl--Sommerfeld expansion}
\label{sec:weyl}

The cylindrical harmonics admit the following plane-wave angular spectrum in the half-space $y<0$ below the duct axis (recall that $\theta$ is measured from the $+y$ axis):
\begin{equation}
H_m^{(1)}(kr)\,\ee^{\ii m\theta}=\frac{\ii^m}{\piup}\int_{-\infty}^{+\infty}
\frac{1}{\gamma(\nu)}\left(\frac{\gamma(\nu)-\ii\nu}{k}\right)^{\!m}
\ee^{\ii\nu z-\ii\gamma(\nu)y}\,\dd\nu,\qquad y<0,
\label{eq:weyl}
\end{equation}
\noindent where $\gamma(\nu)=\sqrt{k^2-\nu^2}$ with $\mathrm{Im}\,\gamma\ge0$; for $|\nu|>k$ one has $\gamma=\ii\sqrt{\nu^2-k^2}$ and the component decays exponentially towards the plate. The $m$-dependent factor is the {algebraic} factor $((\gamma-\ii\nu)/k)^m$ (equivalently $k^m/(\gamma+\ii\nu)^m$, since $\gamma^2+\nu^2=k^2$): on the propagating band $|\nu|\le k$ it equals $\mathrm{exp}({-\ii m\vartheta})$ with $\vartheta=\arcsin(\nu/k)$, while on the evanescent band $|\nu|>k$ it decays algebraically as $(k/2|\nu|)^m$. An earlier version of this work used the unit-modulus factor $\mathrm{exp}({-\ii m\arg(\gamma+\ii\nu)})$ instead; that factor fails to decay in the evanescent band and produced errors of about $10\%$ at $m=5$, which disappeared when the algebraic form was adopted (see \S\ref{sec:numerics}). The identity holds for complex $k$ with $\mathrm{Im}\,k\ge0$ by analytic continuation.

{A few remarks on (\ref{eq:weyl}) are necessary. For $m=0$ it reduces to the classical Sommerfeld identity (\ref{eq:sommerfeld0}), which expresses a two-dimensional cylindrical wave as a superposition of plane waves. The extension to $m>0$ is obtained by applying the ladder operator (\ref{eq:ladderpolar}) $m$ times, as shown below in Steps 1--4. The algebraic factor $((\gamma-\mathrm{i}\nu)/k)^m$ in (\ref{eq:weyl}) is essential: it reduces to the pure phase $\mathrm{e}^{-\mathrm{i}m\theta}$ on the propagating branch ($|\nu|\le k$) and decays algebraically as $(k/2|\nu|)^m$ on the evanescent branch ($|\nu|>k$), thereby correctly representing the evanescent near-field of the spinning mode. This decay property is critical for the spectral accuracy reported in \S\ref{sec:numerics}. Although the three-dimensional Weyl identity for a spherical wave is often quoted alongside the Sommerfeld identity, the present derivation requires only the cylindrical (two-dimensional) form (\ref{eq:weyl}); the term ``Weyl--Sommerfeld'' is used here in the broad sense to denote this class of plane-wave angular-spectrum representations.}
The next derivation proceeds in four steps. 

\emph{Step 1: the $m=0$ Sommerfeld identity.}  For $m=0$, (\ref{eq:weyl}) is the standard Sommerfeld identity
\begin{equation}
H_0^{(1)}(kr)=\frac{1}{\piup}\int_{-\infty}^{+\infty}\frac{1}{\gamma(\nu)}\,
\ee^{\ii\nu z-\ii\gamma(\nu) y}\,\dd\nu,\qquad y<0,
\label{eq:sommerfeld0}
\end{equation}
\noindent a plane-wave (angular) spectrum of the outgoing cylindrical wave: each component satisfies the two-dimensional Helmholtz equation in $(z,y)$, the propagating components $|\nu|\le k$ travel obliquely towards $y<0$, and the evanescent components $|\nu|>k$ decay exponentially away from the axis, ensuring convergence.

\emph{Step 2: the ladder operator in polar form.}  Regard $(z,y)$ as plane Cartesian coordinates and measure the polar angle $\theta'$ from the $+z$ axis, so that $z=r\cos\theta'$, $y=r\sin\theta'$ and $\theta'=\piup/2-\theta$.  The chain rule gives
$\partial_z=\cos\theta'\,\partial_r-(\sin\theta'/r)\partial_{\theta'}$ and
$\partial_y=\sin\theta'\,\partial_r+(\cos\theta'/r)\partial_{\theta'}$,
hence the ladder operator
\begin{equation}
\mathcal O^*=\partial_z-\ii\partial_y
=\ee^{-\ii\theta'}\left(\partial_r-\frac{\ii}{r}\,\partial_{\theta'}\right).
\label{eq:ladderpolar}
\end{equation}
Here \(\theta'\) is an auxiliary angle (from \(+z\)) used only for the ladder-operator step; it connects to the physical angle \(\theta\) (from \(+y\)) by \(\theta' = \pi/2 - \theta\).

\emph{Step 3: induction on the azimuthal order.}  Acting with (\ref{eq:ladderpolar}) on a cylindrical harmonic $Z_m(kr)\mathrm{exp}({-\ii m\theta'})$ and using the cylinder-function recurrence $kZ_m'(kr)-(m/r)Z_m(kr)=-kZ_{m+1}(kr)$ \citep{watson1966},
\begin{equation}
\mathcal O^*\!\left[Z_m(kr)\,\ee^{-\ii m\theta'}\right]
=\ee^{-\ii\theta'}\left(kZ_m'(kr)-\frac{m}{r}Z_m(kr)\right)\ee^{-\ii m\theta'}
=-k\,Z_{m+1}(kr)\,\ee^{-\ii(m+1)\theta'} .
\label{eq:ladderstep}
\end{equation}
Iterating (\ref{eq:ladderstep}) from $m=0$ gives
\begin{equation}
(\mathcal O^*)^m H_0^{(1)}(kr)=(-k)^m H_m^{(1)}(kr)\,\ee^{-\ii m\theta'} .
\label{eq:ladder}
\end{equation}

\emph{Step 4: spectral action and conversion of the angle.}  On the $y<0$ side each component of (\ref{eq:sommerfeld0}) is $\ee^{\ii\nu z-\ii\gamma y}$, so that $\partial_z\to\ii\nu$ and $\partial_y\to-\ii\gamma$, and therefore
\begin{equation}
\mathcal O^*\to\ii\nu-\ii(-\ii\gamma)=\ii\nu-\gamma,
\qquad
(\mathcal O^*)^m\to(\ii\nu-\gamma)^m .
\label{eq:ladderspectral}
\end{equation}
Applying (\ref{eq:ladder}) to (\ref{eq:sommerfeld0}) under the integral and dividing by $(-k)^m$,
\begin{equation}
H_m^{(1)}(kr)\,\ee^{-\ii m\theta'}
=\frac{1}{\piup}\int_{-\infty}^{+\infty}\frac{1}{\gamma}
\left(\frac{\ii\nu-\gamma}{-k}\right)^{\!m}\ee^{\ii\nu z-\ii\gamma y}\,\dd\nu
=\frac{1}{\piup}\int_{-\infty}^{+\infty}\frac{1}{\gamma}
\left(\frac{\gamma-\ii\nu}{k}\right)^{\!m}\ee^{\ii\nu z-\ii\gamma y}\,\dd\nu .
\label{eq:weyltheta}
\end{equation}
Finally $\mathrm{exp}({\ii m\theta})=\mathrm{exp}({\ii m(\piup/2-\theta')})=\ii^m\mathrm{exp}({-\ii m\theta'})$, and multiplying (\ref{eq:weyltheta}) by $\ii^m$ yields (\ref{eq:weyl}). 

Physically, (\ref{eq:weyl}) is the explicit carrier of the change of coordinates from the axisymmetric system of the duct to the Cartesian system of the plate. The azimuthal order $m$ of the spinning mode fixes, through the algebraic factor $((\gamma_d-\ii\nu)/(\lambda_0\omega))^m$, the centroid and the phase structure of the spanwise spectrum of the incident field. Components with $|\nu|>\lambda_0\omega$ are evanescent and largely decay before reaching the plate, but they are still diffracted by the edge. The current spectral framework includes them automatically.

\subsubsection{Double spectrum of the incident normal velocity}
\label{sec:incspectrum}

The rigid-plate boundary condition requires the incident normal velocity $v_i=\partial_y\psi_d$ (the plate normal being the $y$ direction). The assembly of $\widehat A(u,\nu)$ is entirely explicit. Substituting (\ref{eq:weyl}) with $k=\lambda_0(u)\omega$ into (\ref{eq:onplate}) expresses $v_i$ as a double integral over $(u,\nu)$; the normal derivative acts component by component as
\begin{equation}
\partial_y\,\ee^{\ii\nu z-\ii\gamma_d y}=-\ii\gamma_d\,\ee^{\ii\nu z-\ii\gamma_d y},
\end{equation}
and evaluation on the plate $y=-d$ turns the exponential into $\mathrm{exp}({+\ii\gamma_d d})$. Collecting the prefactors -- $\omega/(2\piup)$ from (\ref{eq:onplate}), $-\ii F_+(u)/[\lambda_0(u)H_m^{(1)\prime}(\lambda_0\omega)]$ from (\ref{eq:onplate}), $\ii^m/[\piup\gamma_d]$ from (\ref{eq:weyl}) and $-\ii\gamma_d$ from the differentiation -- the two $\gamma_d$ factors cancel, giving
\begin{equation}
\frac{\omega}{2\piup}\cdot\frac{\ii^m}{\piup\gamma_d}\cdot(-\ii\gamma_d)\cdot(-\ii)
=-\frac{\omega\,\ii^m}{2\piup^2},
\end{equation}
so that
\begin{equation}
v_i(x,z)=\int_\Gamma\!\!\int_{-\infty}^{+\infty}
\widehat A(u,\nu)\,\ee^{\ii\omega ux+\ii\nu z}\,\dd u\,\dd\nu,
\label{eq:vi}
\end{equation}
\begin{equation}
\widehat A(u,\nu)=
-\frac{\omega\,\ii^m}{2\piup^2}\,
\frac{F_+(u)}{\lambda_0(u)\,H_m^{(1)\prime}(\lambda_0\omega)}
\left(\frac{\gamma_d(\nu)-\ii\nu}{\lambda_0(u)\,\omega}\right)^{\!m}
\ee^{\,\ii\gamma_d(\nu)d+\ii\omega uL},
\label{eq:Ahat}
\end{equation}
\noindent where $\gamma_d(\nu)=\sqrt{(\lambda_0(u)\omega)^2-\nu^2}$.
Each $(u,\nu)$ component is a normal-velocity gust $A\,\ee^{\ii k_x x+\ii\nu z}$ on the plate with $k_x=\omega u$, $A=\widehat A(u,\nu)$, and wall-normal wavenumber $\gamma_d$: a (possibly evanescent) plane-wave component of the three-dimensional Helmholtz equation. 

Indeed, since the ambient is quiescent and $\lambda_0(u)=\sqrt{1-u^2}$ (\S\ref{sec:stage1field}),
\begin{equation}
k_x^2+\gamma_d^2+\nu^2
=\omega^2u^2+\omega^2(1-u^2)-\nu^2+\nu^2
=\omega^2,
\label{eq:helmcheck}
\end{equation}
so every spectral component individually satisfies the Helmholtz equation with wavenumber $\omega$, as it must. No expansion coefficient is fitted: every factor in (\ref{eq:Ahat}) traces back to the stage-1 solution through $F_+(u)$, and the constants assembled in the companion code agree with (\ref{eq:Ahat}) to machine precision.

\subsection{Stage 2: scattering of each spectral component by the plate}
\label{sec:stage2}

\subsubsection{Statement of the problem}
\label{sec:component}

Fix $(u,\nu)$ and consider the incident gust $v_g=A\mathrm{exp}({\ii k_x x+\ii\nu z})$ on the plate ($x\le0$), with $k_x=\omega u$ and $A=\widehat A(u,\nu)$. Since the governing equation has constant coefficients in $z$, the scattered field inherits the gust's spanwise dependence:
\begin{equation}
\psi_s(x,\tilde y,z)=\psi_s(x,\tilde y)\,\ee^{\ii\nu z}.
\label{eq:zansatz}
\end{equation}
\noindent where $\tilde y= y +d$, with $d$ the distance between the plate and the axis of the duct. 
Substituting (\ref{eq:zansatz}) into the three-dimensional Helmholtz equation $(\partial_x^2+\partial_{\tilde y}^2+\partial_z^2+\omega^2)\psi_s=0$ and using $\partial_z^2\to-\nu^2$ gives the two-dimensional Helmholtz equation in the $(x,\tilde y)$ plane,
\begin{equation}
\left(\partial_x^2+\partial_{\tilde y}^2+\khat^2\right)\psi_s(x,\tilde y)=0,
\qquad
\khat=\sqrt{\omega^2-\nu^2}\qquad(\mathrm{Im}\,\khat\ge0),
\label{eq:khat}
\end{equation}
\noindent with the causal branch selected by giving $\omega$ a small positive imaginary part (the limiting-absorption convention, suppressed in what follows). Because the plate has vanishing thickness and is rigid, the scattered pressure is an {odd} function of $\tilde y$ -- the classical condition for this geometry \citep{noble1958}: the plate can only support a pressure jump, and any even part would have vanishing normal derivative on both faces and hence decouple from the problem. Oddness restricts the analysis to the half-plane $\tilde y\ge0$ with boundary and edge conditions
\begin{equation}
\partial_{\tilde y}\psi_s=-A\,\ee^{\ii k_x x}\quad(x\le0,\ \tilde y=0);\quad
\psi_s=0\quad(x>0,\ \tilde y=0);\quad
\psi_s=O(x^{1/2})\ \ (x\to0^+),
\label{eq:bc2}
\end{equation}
\noindent namely cancellation of the incident normal velocity on the plate, vanishing of the odd field on the prolongation of the plate, and the edge condition expressing that the rigid leading edge supports at most an integrable pressure singularity (equivalently, finite energy in any neighbourhood of the edge). A radiation condition as $\tilde y^2+x^2\to\infty$ closes the problem.

\subsubsection{Transform and Wiener--Hopf equation}
\label{sec:Wiener--Hopf2}

Apply the $x$-wise Fourier transform with kernel $\mathrm{exp}({\ii\alpha x})$ (following the convention of \citet{huang2017}),
\begin{equation}
\beta(\alpha,\tilde y)=\int_{-\infty}^{+\infty}\psi_s(x,\tilde y)\,\ee^{\ii\alpha x}\dd x,
\qquad
\psi_s(x,\tilde y)=\frac{1}{2\piup}\int\beta(\alpha,\tilde y)\,\ee^{-\ii\alpha x}\dd\alpha .
\label{eq:fwd2}
\end{equation}
Two integrations by parts (the boundary terms vanish by the radiation condition) show that differentiation maps to multiplication by $-\ii\alpha$: $\partial_x^2\to(-\ii\alpha)^2=-\alpha^2$. The Helmholtz equation (\ref{eq:khat}) therefore becomes the ordinary differential equation
\begin{equation}
\partial_{\tilde y}^2\beta=(\alpha^2-\khat^2)\,\beta=\gamma_p^2(\alpha)\,\beta,
\qquad
\gamma_p(\alpha)=\sqrt{\alpha^2-\khat^2},
\label{eq:gammap}
\end{equation}
with $\mathrm{Re}\,\gamma_p\ge0$ on the real $\alpha$ axis (the causal sheet, selected by $\mathrm{Im}\,\khat>0$). The solution satisfying the radiation condition as $|\tilde y|\to\infty$ is
\begin{equation}
\beta(\alpha,\tilde y)=\pm\beta(\alpha,0^\pm)\,\ee^{-\gamma_p(\alpha)|\tilde y|}
\qquad(\tilde y\gtrless0),
\label{eq:radsol2}
\end{equation}
the sign following from the oddness of $\psi_s$ in $\tilde y$. Differentiating (\ref{eq:radsol2}) at $\tilde y=0^+$ gives the relation
\begin{equation}
\gamma_p\,\beta(\alpha,0^+)+\beta'(\alpha,0^+)=0
\label{eq:halfeq}
\end{equation}
between the two unknown boundary values on the plate plane.

The half-range decomposition $\beta=\beta_-+\beta_+$ with
\begin{equation}
\beta_-(\alpha,\tilde y)=\int_{-\infty}^{0}\psi_s\,\ee^{\ii\alpha x}\dd x,
\qquad
\beta_+(\alpha,\tilde y)=\int_{0}^{+\infty}\psi_s\,\ee^{\ii\alpha x}\dd x
\label{eq:halfrange}
\end{equation}
defines functions analytic in overlapping half-planes: since $\psi_s$ grows at most algebraically and radiates, $\beta_-$ is analytic for $\mathrm{Im}\,\alpha<\mathrm{Im}\,\khat$ (a lower half-plane) and $\beta_+$ for $\mathrm{Im}\,\alpha>-\mathrm{Im}\,\khat$ (an upper half-plane); with $\mathrm{Im}\,\khat>0$ the two overlap in the strip $|\mathrm{Im}\,\alpha|<\mathrm{Im}\,\khat$ containing the real axis, which is what makes the Wiener--Hopf algebra below well defined.

The boundary conditions (\ref{eq:bc2}) now translate as follows. First, $\psi_s=0$ on $x>0$ gives $\beta_+(\alpha,0^+)=0$, while $\beta_-(\alpha,0^+)$ remains unknown. Secondly, the plate condition fixes the transformed normal derivative on $x<0$: assuming provisionally $\mathrm{Im}(\alpha+k_x)>0$ for convergence,
\begin{align}
\beta_-'(\alpha,0^+)
&=-A\int_{-\infty}^0 \ee^{\ii(\alpha+k_x)x}\dd x
=-A\left[\frac{\ee^{\ii(\alpha+k_x)x}}{\ii(\alpha+k_x)}\right]_{-\infty}^{0}
\nonumber\\
&=-\frac{A}{\ii(\alpha+k_x)}
=\frac{\ii A}{\alpha-\alpha_p},
\qquad \alpha_p\equiv-k_x=-\omega u ,
\label{eq:betaprime}
\end{align}
the result extending by analytic continuation to the whole strip. Substitution of $\beta_+=0$ at $\tilde y=0^+$ and (\ref{eq:betaprime}) into (\ref{eq:halfeq}) -- written for the sum $\beta=\beta_-+\beta_+$ as
$\gamma_p\beta_-+\beta_-'+\gamma_p\beta_++\beta_+'=0$ -- yields the scalar Wiener--Hopf equation
\begin{equation}
\gamma_p(\alpha)\,\beta_-(\alpha)+\beta_+'(\alpha)=G(\alpha)
\equiv-\frac{\ii A}{\alpha-\alpha_p},
\label{eq:Wiener--Hopfeq2}
\end{equation}
in which $\beta_-$ is analytic in the lower half-plane, $\beta_+'$ in the upper half-plane, and the equation holds in the overlap strip.

The pole position mirrors that of the problem treated by \citet{huang2017}: along the stage-one contour $\Gamma$ (indented downwards, $\mathrm{Im}\,u\le0$) one has $\mathrm{Im}\,\alpha_p=-\mathrm{Im}(\omega u)\ge0$, so the pole lies just \emph{above} the real axis; the inversion contour must pass \emph{below} $\alpha_p$. (On contour segments where $\mathrm{Im}\,u>0$ the pole lies below the axis and the same rule applies unchanged: the $\alpha$ contour passes below $\alpha_p$. The residue correction this requires is derived in \S\ref{sec:rescorr}.)

\subsubsection{Factorisation and solution}
\label{sec:solution2}

The kernel splits {explicitly},
\begin{equation}
\gamma_p=\gamma_+\gamma_-,\qquad \gamma_\pm(\alpha)=\sqrt{\alpha\pm\khat},
\label{eq:split}
\end{equation}
\noindent and no Cauchy integral is needed at this stage. The branch placements are fixed by causality: with $\mathrm{Im}\,\khat>0$, the branch point $-\khat$ of $\gamma_+$ lies in the lower half-plane and its cut is taken downwards, so $\gamma_+$ is analytic and non-zero in the upper half-plane $R^+$; symmetrically, the branch point $+\khat$ of $\gamma_-$ lies in the upper half-plane with its cut upwards, so $\gamma_-$ is analytic and non-zero in the lower half-plane $R^-$. Both are analytic in the strip.

Dividing (\ref{eq:Wiener--Hopfeq2}) by $\gamma_+$,
\begin{equation}
\gamma_-(\alpha)\,\beta_-(\alpha)+\frac{\beta_+'(\alpha)}{\gamma_+(\alpha)}
=-\frac{\ii A\,\gamma_+^{-1}(\alpha)}{\alpha-\alpha_p}.
\label{eq:Wiener--Hopfdiv}
\end{equation}
The only obstruction to analyticity on the right-hand side is the simple pole at $\alpha_p\in R^+$; removing its residue by the additive split
\begin{equation}
\frac{\gamma_+^{-1}(\alpha)}{\alpha-\alpha_p}
=\underbrace{\frac{\gamma_+^{-1}(\alpha)-\gamma_+^{-1}(\alpha_p)}{\alpha-\alpha_p}}_{\text{analytic in }R^+}
+\underbrace{\frac{\gamma_+^{-1}(\alpha_p)}{\alpha-\alpha_p}}_{\text{analytic in }R^-}
\label{eq:addsplit}
\end{equation}
and rearranging gives
\begin{equation}
\underbrace{\gamma_-(\alpha)\,\beta_-(\alpha)
+\frac{\ii A\,\gamma_+^{-1}(\alpha_p)}{\alpha-\alpha_p}}_{\text{analytic in }R^-}
=
\underbrace{-\frac{\beta_+'(\alpha)}{\gamma_+(\alpha)}
-\ii A\,\frac{\gamma_+^{-1}(\alpha)-\gamma_+^{-1}(\alpha_p)}{\alpha-\alpha_p}}_{\text{analytic in }R^+}
=E(\alpha),
\label{eq:Wiener--Hopfrearr}
\end{equation}
where the two sides, analytic in overlapping half-planes and equal in the strip, define a single entire function $E(\alpha)$ by analytic continuation.

It remains to bound $E$ at infinity. The algebraic factors behave as $\gamma_\pm=O(|\alpha|^{1/2})$. The edge condition $\psi_s=O(x^{1/2})$ as $x\to0^+$ translates, by the standard Abelian--Tauberian correspondence for half-range transforms, into $\beta_-(\alpha)=O(|\alpha|^{-3/2})$ as $\alpha\to\infty$ in $R^-$, while $\beta_+'(\alpha)=O(|\alpha|^{-1})$ in $R^+$ (the transform of an integrable function with a bounded edge limit). Both sides of (\ref{eq:Wiener--Hopfrearr}) are therefore $O(|\alpha|^{-1})$ at infinity, so $E(\alpha)\to0$; by Liouville's theorem $E\equiv0$. Solving the left-hand side for $\beta_-$,
\begin{equation}
\beta_-(\alpha;\,u,\nu)=
\frac{-\ii A\;\gamma_+^{-1}(\alpha_p)}{\gamma_-(\alpha)\,(\alpha-\alpha_p)}
=
\frac{-\ii\,\widehat A(u,\nu)}
{\gamma_+(-\omega u)\;\gamma_-(\alpha)\;(\alpha+\omega u)},
\qquad A=\widehat A(u,\nu).
\label{eq:betasol}
\end{equation}
This has the same structure as the straight-edge solution of \citet[eq.~(2.40)]{huang2017}, $\beta_-^{JFM}=G_0^{J}\gamma_+^{-1}(k\cos\theta)/[\gamma_-(\alpha)(\alpha-k\cos\theta)]$ with $G_0^{J}=-k\sin\theta\,\Upsilon$, the only difference being the mirrored pole, which originates from the opposite incidence-direction convention.

A useful analytic check is available. The transformed normal derivative on the plate plane is
\begin{equation}
\beta'(\alpha,0^+)=-\gamma_p(\alpha)\,\beta_-(\alpha)
=\frac{\ii A\,\gamma_+(\alpha)}{\gamma_+(\alpha_p)\,(\alpha-\alpha_p)},
\label{eq:bcheck}
\end{equation}
in which $\gamma_-$ cancels exactly, leaving only the pole $\alpha_p$ in the upper half-plane. In the inverse transform (\ref{eq:fwd2}) with $x<0$, the factor $\ee^{-\ii\alpha x}$ decays in the upper half-plane; closing the contour upwards picks up the residue $\ii A\,\gamma_+(\alpha_p)/\gamma_+(\alpha_p)=\ii A$, and
\begin{equation}
\partial_{\tilde y}\psi_s\big|_{\tilde y=0^+}
=\frac{1}{2\piup}\,(2\piup\ii)\,(\ii A)\,\ee^{-\ii\alpha_p x}
=-A\,\ee^{\ii k_x x}\qquad(x<0),
\label{eq:bcheckres}
\end{equation}
i.e.\ the boundary condition (\ref{eq:bc2}) is recovered identically. 

\subsubsection{The scattered field and its physical decomposition}
\label{sec:scatfield}

Superposing the components,
\begin{equation}
\psi_s(x,\tilde y,z)=\sgn(\tilde y)
\int_\Gamma\!\!\dd u\int \dd\nu\;\ee^{\ii\nu z}\,
\frac{1}{2\piup}\int_{C_\alpha}
\beta_-(\alpha;u,\nu)\,\ee^{-\gamma_p(\alpha)|\tilde y|}\,\ee^{-\ii\alpha x}\,\dd\alpha ,
\label{eq:scatfield}
\end{equation}
with $p_s=\ii\omega\psi_s$. The total field is the sum of the duct free field and the scattered field, $p=p_d+p_s$, where $p_d$ is given by (\ref{eq:ductfield}) at any field point.

The $\alpha$ inversion has the classical Sommerfeld structure: the residue at the pole $\alpha_p$ is the \emph{geometrical reflection} wave and the branch cut of $\gamma_-$ gives the \emph{edge diffraction} field. Both contributions are identified explicitly. From (\ref{eq:betasol}), the residue of the full inversion integrand at $\alpha=\alpha_p$ is
\begin{equation}
\operatorname*{Res}_{\alpha=\alpha_p}
\beta_-(\alpha)\,\ee^{-\gamma_p(\alpha)|\tilde y|-\ii\alpha x}
=-\frac{\ii A}{\gamma_p(\alpha_p)}\,
\ee^{-\gamma_p(\alpha_p)|\tilde y|+\ii k_x x}.
\label{eq:respole}
\end{equation}
The square root at the pole is evaluated on the causal sheet as follows. For a propagating component ($|k_x|<\khat$) with $\mathrm{Im}\,\khat>0$, the point $\alpha_p+\khat=\khat-k_x$ approaches the positive real axis from above, so $\gamma_+(\alpha_p)=+\sqrt{\khat-k_x}$; the point $\alpha_p-\khat=-(\khat+k_x)$ approaches the negative real axis from below, so its argument is $-\piup$ and $\gamma_-(\alpha_p)=\ee^{-\ii\piup/2}\sqrt{\khat+k_x}=-\ii\sqrt{\khat+k_x}$. Hence
\begin{equation}
\gamma_p(\alpha_p)=\gamma_+(\alpha_p)\gamma_-(\alpha_p)
=-\ii\sqrt{\khat^2-k_x^2}=-\ii k_y,
\qquad k_y\equiv\sqrt{\khat^2-k_x^2}.
\label{eq:gpalphap}
\end{equation}
For an observer with $x<0$ the factor $\mathrm{exp}({-\ii\alpha x})$ decays in the upper half-plane; closing the (below-pole) contour upwards encircles $\alpha_p$ counter-clockwise, and (\ref{eq:respole})--(\ref{eq:gpalphap}) give
\begin{equation}
\psi_s^{\rm geom}=\frac{1}{2\piup}\,(2\piup\ii)\left(-\frac{\ii A}{\gamma_p(\alpha_p)}\right)
\ee^{\ii k_x x+\ii k_y\tilde y}
=\frac{\ii A}{k_y}\,\ee^{\ii k_x x+\ii k_y\tilde y},
\qquad \tilde y>0,
\label{eq:geom}
\end{equation}
\noindent in the reflection zone. The reflection zone itself follows from ray geometry: tracing the ray from the observer $(\rho\cos\varphi,\rho\sin\varphi)$ back along the reflected direction $-(k_x,k_y)$, it reaches the plate plane $\tilde y=0$ after a distance parameter $t_0=\rho\sin\varphi/k_y$, landing at $x_0=\rho\cos\varphi-(k_x/k_y)\rho\sin\varphi$; the ray has actually reflected on the plate if and only if $x_0\le0$, i.e.
\begin{equation}
\text{reflection zone: }\ \cos\varphi\le \frac{k_x}{k_y}\sin\varphi,
\label{eq:refzone}
\end{equation}
where $\varphi$ is the polar angle of the observation direction in the $(x,\tilde y)$ plane measured from the $+x$ axis. Evanescent components ($k_y$ imaginary, $\gamma_p(\alpha_p)>0$) produce no geometrical reflection, only a near-field evanescent contribution decaying as $\mathrm{exp}({-\gamma_p(\alpha_p)|\tilde y|})$, and edge diffraction.

\subsubsection{Pole crossing along the deformed contour}
\label{sec:rescorr}

The rule of \S\ref{sec:Wiener--Hopf2} -- the $\alpha$ contour $C$ must pass \emph{below} $\alpha_p$ -- holds uniformly along the stage-one contour $\Gamma$. On the segments of $\Gamma$ that rise above the real axis ($\mathrm{Im}\,u>0$, the upstream-propagating part of the deformed contour), however, $\alpha_p=-\omega u$ has $\mathrm{Im}\,\alpha_p=-\mathrm{Im}(\omega u)<0$: the pole lies \emph{below} the real axis, so a real-axis computational grid $R$ passes above the pole while the correct contour $C$ passes below it. The closed loop formed by $R$ traversed left-to-right and $C$ traversed right-to-left encircles $\alpha_p$ clockwise, hence
\begin{equation}
\oint_{R\cup(-C)}=-2\piup\ii\,\operatorname*{Res}_{\alpha_p}
\quad\Longrightarrow\quad
\int_C=\int_R+2\piup\ii\,\operatorname*{Res}_{\alpha_p}.
\label{eq:contourdef}
\end{equation}
Dividing by $2\piup$ and inserting (\ref{eq:respole}), the exact correction to the real-axis evaluation is, per crossed component,
\begin{equation}
\Delta\psi_s^{(\alpha)}=\frac{1}{2\piup}\,(2\piup\ii)
\left(-\frac{\ii A}{\gamma_p(\alpha_p)}\right)
\ee^{-\gamma_p(\alpha_p)|\tilde y|+\ii k_x x}
=\frac{A}{\gamma_p(\alpha_p)}\,
\ee^{-\gamma_p(\alpha_p)|\tilde y|+\ii k_x x},
\label{eq:rescorr1}
\end{equation}
Restoring the spanwise superposition and the $u$ quadrature along $\Gamma$ (with $w_u$ the local quadrature weight of the crossed sample), the correction implemented in the code is
\begin{equation}
\Delta\psi_s=\sgn(\tilde y)\,w_u\int \dd\nu\,
\frac{\widehat A}{\gamma_p(\alpha_p)}\,
\ee^{-\gamma_p(\alpha_p)|\tilde y|+\ii k_x x+\ii\nu z},
\label{eq:rescorr}
\end{equation}
with $\gamma_p(\alpha_p)=\gamma_+(\alpha_p)\gamma_-(\alpha_p)$ evaluated on the causal sheet (\ref{eq:gpalphap}). For a propagating component (\ref{eq:rescorr}) is precisely the geometrical-reflection plane wave (\ref{eq:geom}): the correction is nothing but the analytic continuation of the reflection wave, which the naive real-axis integral loses when the pole crosses to the other side. Its effect on the solution accuracy is quantified in \S\ref{sec:numerics}.

\subsection{Far-field asymptotics}
\label{sec:farfield}

\subsubsection{Saddle point of the $\alpha$ integral}
\label{sec:saddlealpha}

For an observer at $(x,\tilde y)=(\rho\cos\varphi,\rho\sin\varphi)$ with $\khat\rho\gg1$, consider the $\alpha$ inversion of a single $(u,\nu)$ component. Introduce the standard angular parametrisation of the hyperbolic contour,
\begin{equation}
\alpha=-\khat\cos\eta,
\qquad
\gamma_p=\sqrt{\alpha^2-\khat^2}=-\ii\khat\sin\eta,
\qquad
\dd\alpha=\khat\sin\eta\,\dd\eta,
\label{eq:alphaparam}
\end{equation}
where the branch $\gamma_p=-\ii\khat\sin\eta$ maps the real $\alpha$ axis (with the causal displacement $\mathrm{Im}\,\khat>0$) onto a steepest-descent path through $\eta=\varphi$. The phase of the integrand becomes, using $|x|=-x$ for $x<0$ or directly for the present geometry,
\begin{equation}
-\gamma_p(\alpha)|\tilde y|-\ii\alpha x
=\ii\khat\sin\eta\;\rho\sin\varphi+\ii\khat\cos\eta\;\rho\cos\varphi
=\ii\khat\rho\cos(\eta-\varphi)
\equiv\ii\Phi(\eta).
\label{eq:alphaphase}
\end{equation}
The saddle is at $\Phi'(\eta)=-\khat\rho\sin(\eta-\varphi)=0$, i.e.\ $\eta_s=\varphi$, at which
\begin{equation}
\alpha_s=-\khat\cos\varphi,
\qquad
\gamma_p(\alpha_s)=-\ii\khat\sin\varphi,
\qquad
\Phi(\varphi)=\khat\rho,
\qquad
\Phi''(\varphi)=-\khat\rho .
\label{eq:alphasaddle}
\end{equation}
Since $\Phi''(\varphi)$ is real and negative, the stationary-phase formula
$$\int g(\eta)\,\ee^{\ii\Phi}\dd\eta\sim g(\eta_s)\sqrt{2\piup/|\Phi''(\eta_s)|}\,\ee^{\ii\Phi(\eta_s)-\ii\piup/4}$$
applies with $g=\beta_-(\alpha(\eta))\,\khat\sin\eta$, and the factor $\khat\sin\varphi\sqrt{2\piup/(\khat\rho)}$ reduces to $\sin\varphi\sqrt{2\piup\khat/\rho}$; dividing by $2\piup$,
\begin{equation}
\frac{1}{2\piup}\int_{C_\alpha}\beta_-\,\ee^{-\gamma_p|\tilde y|-\ii\alpha x}\dd\alpha
\;\sim\;
\beta_-(\alpha_s)\,\sin\varphi\,\sqrt{\frac{\khat}{2\piup\rho}}\,
\ee^{-\ii\piup/4}\,\ee^{\ii\khat\rho}
\qquad(0<\varphi<\piup),
\label{eq:saddle1}
\end{equation}
to which the geometrical term (\ref{eq:geom}) must be added inside the reflection zone. The cylindrical spreading $\rho^{-1/2}$ is the expected two-dimensional edge-diffraction decay at fixed $\nu$; the three-dimensional spreading is restored by the $\nu$ saddle below.

\subsubsection{The three-dimensional far field}
\label{sec:saddlenu}

Introduce spherical coordinates about the plate edge, $R=\sqrt{x^2+\tilde y^2+z^2}$, with direction $\hat{\boldsymbol n}=(n_x,n_y,n_z)$, so that $x=Rn_x$, $\tilde y=Rn_y$, $z=Rn_z$ and the cylindrical radius in (\ref{eq:saddle1}) is $\rho=R\sqrt{1-n_z^2}$. The $\nu$ integral of each $u$ component has, from (\ref{eq:saddle1}), the phase
\begin{equation}
\Phi(\nu)=\rho\,\khat+\nu z=\rho\sqrt{\omega^2-\nu^2}+\nu z,
\qquad
\Phi'(\nu)=-\frac{\nu\rho}{\khat}+z .
\label{eq:nuphase}
\end{equation}
The saddle condition $\Phi'(\nu_s)=0$ gives $\nu_s\rho/\khat_s=z$, i.e.\ $$\nu_s/\sqrt{\omega^2-\nu_s^2}=n_z/\sqrt{1-n_z^2},$$ 
whose solution is
\begin{equation}
\nu_s=\omega n_z,
\qquad
\khat_s=\sqrt{\omega^2-\nu_s^2}=\omega\sqrt{1-n_z^2},
\qquad
\alpha_s=-\khat_s\cos\varphi=-\omega n_x ,
\label{eq:nusaddle}
\end{equation}
the last equality using $\cos\varphi=x/\rho=n_x/\sqrt{1-n_z^2}$. The second derivative is
\begin{equation}
\Phi''(\nu)=-\frac{\rho}{\khat}-\frac{\nu^2\rho}{\khat^3}
=-\frac{\rho(\khat^2+\nu^2)}{\khat^3}
=-\frac{\rho\,\omega^2}{\khat^3},
\qquad
\Phi''(\nu_s)=-\frac{\rho\,\omega^2}{\khat_s^3}=-\frac{R^3}{\omega\rho^2},
\label{eq:nusecond}
\end{equation}
using $\khat_s=\omega\rho/R$ in the last step. The phase at the saddle is
\begin{equation}
\Phi(\nu_s)=\rho\,\khat_s+\nu_s z
=\frac{\omega\rho^2}{R}+\omega n_z^2 R
=\frac{\omega}{R}\left(\rho^2+n_z^2R^2\right)=\omega R .
\label{eq:nuphaseval}
\end{equation}
Applying stationary phase to the $\nu$ integral contributes the factor
$\sqrt{2\piup/|\Phi''(\nu_s)|}\,\ee^{\ii\omega R-\ii\piup/4}
=\sqrt{2\piup\omega\rho^2/R^3}\,\ee^{\ii\omega R-\ii\piup/4}$; combining with the $\alpha$-saddle amplitude (\ref{eq:saddle1}), the two square roots give
\begin{equation}
\sin\varphi\,\sqrt{\frac{\khat_s}{2\piup\rho}}\cdot\sqrt{\frac{2\piup\omega\rho^2}{R^3}}
=\sin\varphi\,\sqrt{\frac{\khat_s\,\omega\rho}{R^3}}
=\sin\varphi\,\frac{\omega\sqrt{1-n_z^2}}{R}
=\frac{\omega n_y}{R},
\label{eq:faramp}
\end{equation}
where $n_y=\rho\sin\varphi/R=\sin\varphi\sqrt{1-n_z^2}$ has been used, and the two phase factors combine as $\ee^{-\ii\piup/4}\cdot\ee^{-\ii\piup/4}=\ee^{-\ii\piup/2}=-\ii$. The diffracted far field is therefore
\begin{equation}
\psi_s^{\rm diff}(R,\hat{\boldsymbol n})\sim
\mathcal D_\psi(\hat{\boldsymbol n})\,\frac{\ee^{\ii\omega R}}{R},\qquad
\mathcal D_\psi(\hat{\boldsymbol n})=
-\ii\omega\, n_y
\int_\Gamma\beta_-\big(\alpha_s;u,\nu_s\big)\Big|_{\alpha_s=-\omega n_x,\ \nu_s=\omega n_z}
\dd u ,
\label{eq:Dpsi}
\end{equation}
for $n_y>0$, with odd continuation in $n_y$; the constant in (\ref{eq:Dpsi}) is confirmed against a full double numerical quadrature in \S\ref{sec:numerics}. Note the expected spherical spreading $R^{-1}$ and the edge directivity factor $n_y$: the edge radiates like a line source dipole in the plate-normal direction, vanishing in the plate plane. The geometrical-reflection part reaches the far field only along the specular cone (a $\delta$-type contribution in direction space) and is evaluated separately below. The $u$ integral in (\ref{eq:Dpsi}) is left unevaluated: the incident spectrum is broadband and known only along the deformed contour $\Gamma$, so no isolated saddle exists at the frequencies of interest; this single contour integral is the only numerical quadrature surviving in the far field. The diffracted directivity plotted in \S\ref{sec:results} is $\mathcal D_p=\ii\omega\mathcal D_\psi$. Equation (\ref{eq:Dpsi}) is the final output of the coupled model: a complete transfer function from the incident duct mode $(m,\mu)$ to the scattered far-field directivity.

\subsubsection{Geometrical reflection: the image-source identity}
\label{sec:image}

Summing the per-component reflection residues (\ref{eq:geom}) directly along the deformed complex contour $\Gamma$ is ill conditioned: on contour segments with $\mathrm{Im}\,u\ne0$ the individual terms behave as $\mathrm{exp}({\pm\omega\,\mathrm{Im}(u)R})$ and overflow at large $R$ (values of order $10^{114}$ are reached at $R=60$), even though their sum is a well-behaved plane-wave superposition. A strictly equivalent, well-conditioned form is obtained as follows.

The plate is rigid, so each incident plane-wave component of the Weyl--Sommerfeld spectrum (\ref{eq:weyl}) is reflected with reflection coefficient $+1$: the reflected component carries the same amplitude, the same $x$ and $z$ wavenumbers, and the opposite wall-normal wavenumber. But the superposition of all such reflected components is, by construction, nothing but the incident duct field continued across the plate plane as if radiated from the mirror point -- the standard image argument for a rigid plane. Precisely, the geometrical-reflection field equals the free radiation of the duct exit from the image $O_1'=(-L,-2d,0)$ of the exit centre $O_1=(-L,0,0)$ in the plate plane $\tilde y=0$ (the plate coordinate $\tilde y$ being measured from the plate, the image sits at distance $d$ below it).

In the far field this is evaluated by translating the known duct directivity. Let $D_P(\Theta)$ be the free duct-radiation directivity of the first stage, defined by (\ref{eq:ductdir}) with the azimuthal factor $\mathrm{exp}({\ii m\theta})$ extracted. A source point displaced by $\boldsymbol r_0$ from the reference origin contributes, at a distant observer $R\hat{\boldsymbol n}$, the additional phase $\mathrm{exp}({-\ii\omega\hat{\boldsymbol n}\boldsymbol\cdot\boldsymbol r_0})$, since $|\boldsymbol r_0-R\hat{\boldsymbol n}|=R-\hat{\boldsymbol n}\boldsymbol\cdot\boldsymbol r_0+O(R^{-1})$; with $\boldsymbol r_0=O_1'=(-L,-2d,0)$ the image field therefore carries the phase factor $\mathrm{exp}({\ii\omega(Ln_x+2dn_y)})$. The polar angle and azimuth of the observer as seen from the image are $\arccos n_x+O(R^{-1})$ and $\theta_m=\mathrm{atan2}(n_z,-n_y)$ -- note the minus sign on $n_y$, the image lying below the plate plane. Hence
\begin{equation}
 D_{\rm geom}(\hat{\boldsymbol n})=D_P(\arccos n_x)\,\ee^{\ii m\theta_m}\,\ee^{\ii\omega(Ln_x+2d\,n_y)},
\label{eq:image}
\end{equation}
\noindent with
\begin{equation}
\theta_m=\mathrm{atan2}(n_z,-n_y),
\quad
\nonumber  \text{reflection zone: }n_y>0,\quad-L+\frac{d\,n_x}{n_y}<0,
\end{equation}
\noindent with $L=d=2$ in the present geometry; in the $z=0$ plane the azimuthal factor reduces to $\ee^{\ii m\theta_m}=(-1)^m$. The reflection-zone condition in (\ref{eq:image}) is pure ray geometry: the ray from the image point towards $\hat{\boldsymbol n}$ crosses the plate plane $\tilde y=0$ after a path length $s=d/n_y$, landing at $x=-L+(d/n_y)n_x$; the reflection is physical if and only if this landing point lies on the plate half-plane, i.e.\ has $x\le0$. No contour integral over $\Gamma$ is needed for $D_{\rm geom}$: the image identity collapses the whole incident spectrum back onto the stage-one directivity, which is precisely what removes the overflow. The remaining geometrical term -- the incident field itself -- and the final assembly are given in \S\ref{sec:fullcircle}.

\subsubsection{The incident field, the shadow zone and the full-circle directivity}
\label{sec:fullcircle}

Two further elements complete the far field. The first is the incident duct radiation itself, which reaches the observer along the ray from the exit centre $O_1=(-L,0,0)$ unless the ray is blocked by the plate. The ray $O_1+s\hat{\boldsymbol n}$, $s>0$, meets the plate plane $\tilde y=0$ at $s=-d/n_y$, which is positive only for $n_y<0$, and lands at $x_{\rm hit}=-L-d\,n_x/n_y$; since the plate occupies $x\le0$, the incident ray is blocked if and only if
\begin{equation}
n_y<0,
\qquad
x_{\rm hit}=-L-d\,\frac{n_x}{n_y}\le0
\qquad\text{(shadow zone)}.
\label{eq:shadow}
\end{equation}
Within the shadow zone the pole contributions of the scattered field below the plate cancel the incident field exactly -- this is the content of the Wiener--Hopf solution continued to $\tilde y<0$ -- so the geometrical (non-diffracted) field there vanishes identically. As in \S\ref{sec:image}, the residue sum behind this cancellation is not evaluated numerically, the individual terms overflowing at large $R$; the cancellation is imposed through the zone test (\ref{eq:shadow}), which is exact. Outside the shadow zone the incident far field is the duct directivity (\ref{eq:ductdir}) translated from $O_1$ to the plate edge, i.e.\ multiplied by $\mathrm{exp}({-\ii\omega\hat{\boldsymbol n}\boldsymbol\cdot\boldsymbol r_0})$ with $\boldsymbol r_0=(-L,0,0)$:
\begin{equation}
D_{\rm inc}(\hat{\boldsymbol n})=D_P(\arccos n_x)\,\ee^{\ii m\hat\theta}\,\ee^{\ii\omega Ln_x},
\qquad
\hat\theta=\mathrm{atan2}(n_z,n_y),
\qquad
\text{$\hat{\boldsymbol n}$ outside (\ref{eq:shadow})},
\label{eq:Dinc}
\end{equation}
\noindent the azimuthal factor reducing to $\mathrm{exp}({\ii m\hat\theta})=1$ ($n_y>0$) or $(-1)^m$ ($n_y<0$) in the $z=0$ plane.

The second element is a remark on the edge-diffraction term. The scattered field is odd in $\tilde y$ (\S\ref{sec:solution2}), and evaluating the saddle (\ref{eq:alphasaddle}) with $|\tilde y|$ contributes $\mathrm{sgn}(\tilde y)\,|\sin\varphi|=\sin\varphi$:  (\ref{eq:Dpsi}) therefore holds as written on the whole circle of directions, $n_y\gtrless0$, without modification.

The total far-field directivity of the coupled model is finally
\begin{equation}
D_{\rm tot}(\hat{\boldsymbol n})=
\underbrace{\mathcal D_p(\hat{\boldsymbol n})}_{\text{edge diffraction, all }\hat{\boldsymbol n}}
+\underbrace{D_{\rm inc}(\hat{\boldsymbol n})}_{\text{incident, outside (\ref{eq:shadow})}}
+\underbrace{D_{\rm geom}(\hat{\boldsymbol n})}_{\text{reflection zone (\ref{eq:image})}},
\qquad
\mathcal D_p=\ii\omega\mathcal D_\psi ,
\label{eq:Dtot}
\end{equation}
with $\mathcal D_\psi$ from (\ref{eq:Dpsi}). Comparing (\ref{eq:Dtot}) with the duct-alone result (\ref{eq:ductdir}) exposes the action of the plate at the formula level: the duct-alone directivity $D_P$ appears explicitly in two of the three terms, and the plate enters only through (i) the azimuthal flip $\mathrm{exp}({\ii m\theta})\to\mathrm{exp}({\ii m\theta_m})$ and the additional path $2d\,n_y$ of the image term, (ii) the cancellation of the incident term inside the shadow zone (\ref{eq:shadow}), and (iii) the additive edge-diffraction term, which is the only genuinely new directivity. In particular, removing the plate ($\widehat A\to0$, whence $\mathcal D_p\to0$, $D_{\rm geom}\to0$ and the shadow zone empties) reduces (\ref{eq:Dtot}) to $D_{\rm inc}\equiv D_P(\arccos n_x)\,\mathrm{exp}({\ii m\hat\theta})$, i.e.\ exactly to the duct-alone directivity (\ref{eq:ductdir}) of \citet{gabard2006} up to the origin translation $\mathrm{exp}({\ii\omega Ln_x})$ -- a formula-level consistency check of the coupled model. The direction $\varphi=\piup$ in the $z=0$ plane, for which the observer looks along the plate surface, is degenerate for the saddle formula (\ref{eq:Dpsi}) (the factor $\sqrt{\alpha_s-\khat}$ vanishes) and is excluded from the directivity plots of \S\ref{sec:results}.

\subsection{Limiting cases checks}
\label{sec:limits}

Three limiting checks establish the analytical soundness of the coupled model.
\begin{enumerate}
  \item \emph{Plate removed.} As $\widehat A\to0$, the edge and image terms of (\ref{eq:Dtot}) vanish, the shadow zone (\ref{eq:shadow}) empties, and the model degenerates to $D_{\rm tot}=D_{\rm inc}\equiv D_P(\Theta)\,\mathrm{exp}({\ii m\hat\theta})$, the duct-alone far-field directivity of \citet{gabard2006} up to the origin translation (see the end of \S\ref{sec:fullcircle}).
  \item \emph{Single incident plane wave.} Taking $\widehat A$ to be a single propagating component with $k_x=\khat\cos\theta_i$, the identities $\gamma_-(\alpha_s)=-\ii\sqrt{2\khat}\cos(\varphi/2)$, $\gamma_+(-k_x)=\sqrt{\khat(1-\cos\theta_i)}$ and $\alpha_s+k_x=\khat(\cos\theta_i-\cos\varphi)$ reduce (\ref{eq:saddle1})--(\ref{eq:betasol}) to
  \begin{equation}
  |\mathcal D|\propto
  \frac{\sin\varphi}{\cos(\varphi/2)\,\sin(\theta_i/2)\,|\cos\theta_i-\cos\varphi|}
  =\frac{2\sin(\varphi/2)}{\sin(\theta_i/2)\,|\cos\theta_i-\cos\varphi|},
  \label{eq:noble}
  \end{equation}
  the classical rigid half-plane (Sommerfeld/Noble) diffraction directivity \citep{noble1958}.
  \item \emph{High-frequency limit.} A further stationary-phase evaluation of the $u$ integral in (\ref{eq:Dpsi}) locks the saddle onto the duct stationary-phase wavenumber $u_s$ corresponding to the geometrical direction from the exit to the edge, and $\mathcal D$ factorises as $D_P(\Theta_g)\times\mathcal S(\Theta_g\to\hat{\boldsymbol n})$ times a spreading factor: a product of geometrical-acoustics type, which serves as an asymptotic reference for the numerical results.
\end{enumerate}

\section{Numerical implementation and validation}
\label{sec:numerics}

\subsection{Numerical implementation}
\label{sec:implementation}

The implementation is modular and leaves the stage-one code untouched: a preprocessing step extracts $F_+(u)$ from the duct solver and assembles $\widehat A(u,\nu)$ and $\gamma_+^{-1}(-\omega u)$ from (\ref{eq:Ahat}) and (\ref{eq:split}); a field module evaluates (\ref{eq:scatfield}) at arbitrary point sets, accelerating the $\alpha$ inversion (at fixed $\tilde y$) and the $\nu$ synthesis (at fixed $x$) by fast Fourier transforms (FFTs); and a far-field module evaluates (\ref{eq:Dpsi})--(\ref{eq:Dtot}). The $u$ contour reuses the deformed parabolic contour of the duct code (1024 points, which may be trimmed to 256). The $\nu$ integrand has integrable square-root singularities at $\gamma_d=0$ ($|\nu|=\lambda_0\omega$) and at $\gamma_+(\alpha_p)=0$; the propagating band is smoothed by the substitution $\nu=\lambda_0\omega\sin t$ and the evanescent tails are covered by a graded mesh, the window $|\nu|\le1.3\,\omega$ with several hundred points being sufficient. The $\alpha$ grid is uniform over $|\alpha|\le4\omega$ with $N_\alpha\approx512$, and the wavenumber is given a small imaginary part $k=\omega(1+10^{-3}\ii)$ to displace the branch points $\pm\khat$ off the real axis, as in \citet{huang2017}; for near-field evaluation the contour is deformed parabolically towards the side of faster decay according to the sign of $x$.

One implementation subtlety is worth recording. Where the duct contour $\Gamma$ rises above the real axis ($\mathrm{Im}\,u>0$), the incident pole $\alpha_p=-\omega u$ crosses below the real $\alpha$ contour and the rule of \S\ref{sec:Wiener--Hopf2} (the contour passes below $\alpha_p$) requires the exact residue correction (\ref{eq:rescorr}) derived in \S\ref{sec:rescorr}, applied per crossed component with its local quadrature weight (for a propagating component this is precisely the geometrical-reflection plane wave). This correction reduces the plate boundary-condition residual from $O(|v_i|)$ to the percent level of the incident amplitude (\S\ref{sec:validation}). 

The branch enforcement deserves emphasis: without it, contour points with $\mathrm{Im}\,u>0$ and $|\mathrm{Re}\,u|>1$ select the wrong Riemann sheet of the square roots, and the factor $\mathrm{exp}({\ii\gamma_d d})$ in (\ref{eq:Ahat}) then grows exponentially along the contour. After the fixes, the FFT evaluation and a direct double trapezoidal quadrature of the scattered field agree to a complex relative error of $2\times10^{-13}$ (a single point immediately adjacent to the plate shows $1.8\times10^{-7}$), and the exterior field values are $O(10^{-1})$, as expected.

\subsection{Validation}
\label{sec:validation}

Every element of the coupled model was verified against previous numerical solver \citep{huang2008}. For brevity the details regarding the high-oroder spatial and temporal schemes are omitted here. The final status is as follows.
\begin{enumerate}
  \item \emph{Weyl identity} (\ref{eq:weyl}). With the algebraic phase factor, the relative error for $m=0,1,3,5$ is independent of $m$ and converges as $N^{-1/2}$ under evanescent-band mesh refinement -- the expected order for the midpoint rule applied to the integrable $1/\sqrt{\nu\mp k}$ singularities at $\nu=\pm k$; the propagating band is smooth after the substitution $\nu=k\sin t$.
  \item \emph{Wiener--Hopf boundary conditions} (\ref{eq:bc2}). $\psi_s(x>0,\tilde y=0)=0$ holds exactly, and $\partial_{\tilde y}\psi_s=-A\ee^{\ii k_x x}$ for $x<0$ holds to a relative error of $1$--$2\%$, attributable to truncation ringing; recall from \S\ref{sec:solution2} that the identity also holds analytically.
  %
  %
  \item \emph{Three-dimensional far field} (\ref{eq:Dpsi}). The ratio of a full $(\alpha,\nu)$ double quadrature at $R=30$ to the closed-form saddle formula with the geometrical-reflection band included is $1.004$.
  \item \emph{Spectral bridge} (\S\ref{sec:bridge}). For the with-jet configuration of \S\ref{sec:results}, the double spectrum (\ref{eq:vi})--(\ref{eq:Ahat}) reconstructs the incident normal velocity on the plate plane to a relative error of $4.6\times10^{-4}$--$2.1\times10^{-3}$ at eight random points (against a finite-difference evaluation of the duct field), and the incident potential and pressure to $1.0\times10^{-2}$ on the plate plane $y=-2$ and $3.9\times10^{-3}$ on the mid plane $y=-1$; for the quiescent configuration the normal velocity is reconstructed to $1.7$--$2.3\%$, and the potential and pressure to below $5\%$ at most sampled points, with isolated maxima of $13\%$, $7.6\%$ and $3.4\%$ on the planes $y=-2$, $-1$ and $-0.25$ respectively. The residual error is attributed to the truncation of the $u$ window at $u_{cut}=1.3$ and, away from the plate plane, to the re-amplification of the evanescent spectral components truncated at the plate-plane spectral window; the second stage consumes only the plate-plane spectrum.
  \item \emph{End to end.} The FFT conventions agree with direct summation to $2\times10^{-13}$ (one point immediately adjacent to the plate showing $1.8\times10^{-7}$); the total normal velocity $\partial_{\tilde y}\psi_s+v_i$ vanishes on the plate with a residual of $0.13$--$1.3\%$ of the local incident amplitude at random surface points for the quiescent configuration, and $1.6$--$7.6\%$ for the with-jet configuration. The larger residual of the with-jet case is attributed to its richer incident spectrum, more severely affected by the truncation of the $u$ window at $u_{cut}=1.3$. For reference, the earlier quiescent annular-duct validation runs gave a boundary-condition residual of $10^{-3}$--$3\%$ and a total-to-incident surface-pressure ratio of median $1.96$ for $x<-1.5$, consistent with rigid-plate pressure. 
\end{enumerate}

\section{Results}
\label{sec:results}

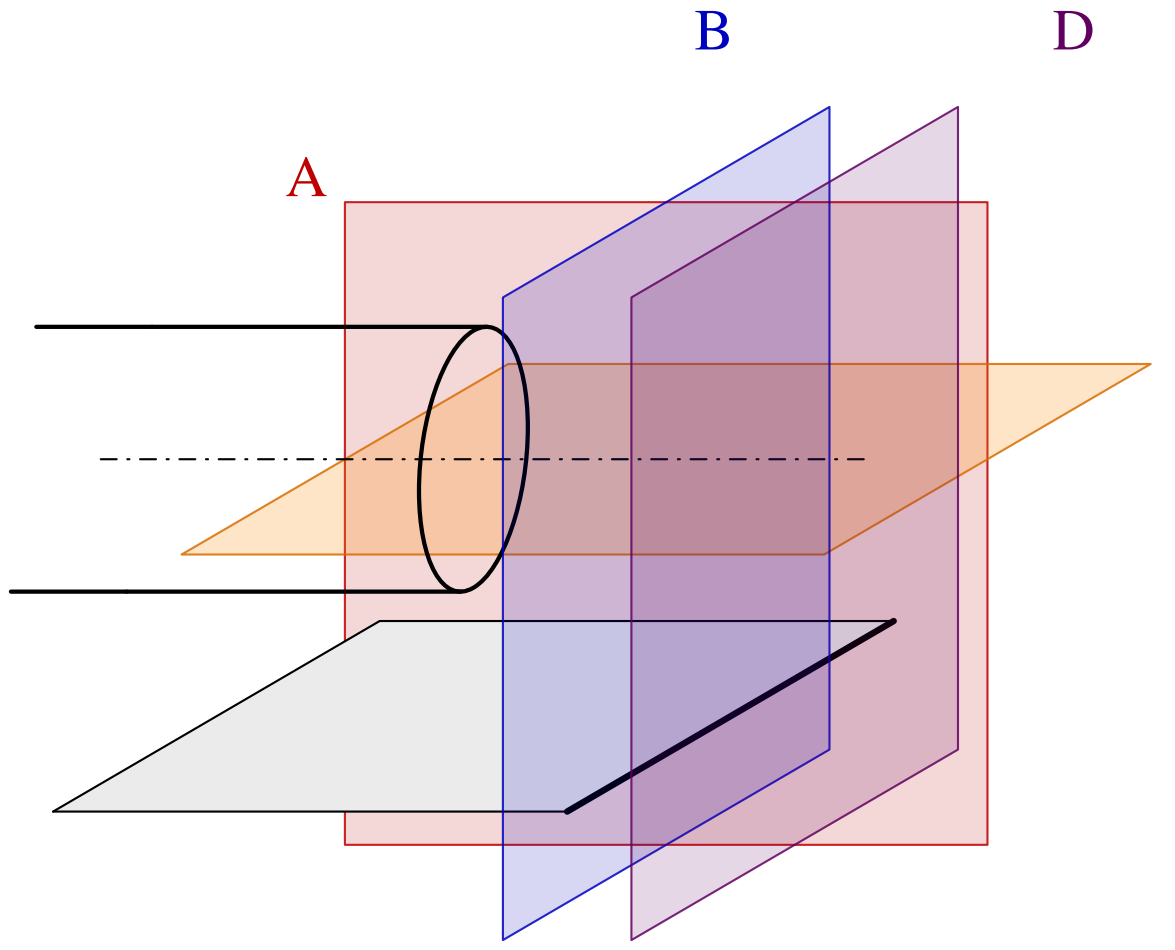
\begin{figure}
  \centering
\begin{tikzpicture}[x={(0.85cm,0cm)}, y={(0cm,0.85cm)}, z={(0.36cm,0.21cm)},
    line join=round, line cap=round, >={Stealth[length=1.6mm,width=1.2mm]}]

  \fill[red!75!black, opacity=0.16] (-3,-3,0) -- (2,-3,0) -- (2,2,0) -- (-3,2,0) -- cycle;
  \draw[red!75!black, opacity=0.85, thin] (-3,-3,0) -- (2,-3,0) -- (2,2,0) -- (-3,2,0) -- cycle;
  \node[red!75!black] at (-3.3,2.2,0) {\large A};

  \fill[orange!90!yellow, opacity=0.22] (-3,0,-3) -- (2,0,-3) -- (2,0,3) -- (-3,0,3) -- cycle;
  \draw[orange!85!black, opacity=0.85, thin] (-3,0,-3) -- (2,0,-3) -- (2,0,3) -- (-3,0,3) -- cycle;
  \node[orange!85!black] at (2.55,0.75,3) {\large C};

  \draw[thick, canvas is yz plane at x=-2] (0,0) circle (1);
  \draw[thick] (-2,0.9724,0.2334) -- (-4.6,0.9724,0.2334);
  \draw[thick] (-2,-0.9724,-0.2334) -- (-4.6,-0.9724,-0.2334);
  \draw[thick] (-4.6,0.9724,0.2334) -- (-5.5,0.9724,0.2334);
  \draw[thick] (-4.6,-0.9724,-0.2334) -- (-5.5,-0.9724,-0.2334);
  \draw[dash dot, thin] (-4.9,0,0) -- (1.1,0,0);

  \fill[gray!16] (-4,-2,-3) -- (-4,-2,3) -- (0,-2,3) -- (0,-2,-3) -- cycle;
  \draw[thin] (-4,-2,-3) -- (0,-2,-3);
  \draw[thin] (-4,-2,3) -- (0,-2,3);
  \draw[thin] (-4,-2,-3) -- (-4,-2,3);
  \draw[very thick] (0,-2,-3) -- (0,-2,3);

  \fill[blue!75!black, opacity=0.16] (-0.5,-3,-3) -- (-0.5,-3,3) -- (-0.5,2,3) -- (-0.5,2,-3) -- cycle;
  \draw[blue!75!black, opacity=0.85, thin] (-0.5,-3,-3) -- (-0.5,-3,3) -- (-0.5,2,3) -- (-0.5,2,-3) -- cycle;
  \node[blue!75!black] at (-1.4,2.6,3) {\large B};

  \fill[violet!75!black, opacity=0.16] (0.5,-3,-3) -- (0.5,-3,3) -- (0.5,2,3) -- (0.5,2,-3) -- cycle;
  \draw[violet!75!black, opacity=0.85, thin] (0.5,-3,-3) -- (0.5,-3,3) -- (0.5,2,3) -- (0.5,2,-3) -- cycle;
  \node[violet!75!black] at (1.4,2.6,3) {\large D};

\end{tikzpicture}
  \caption{Cut planes used to display the near field in the results section: A, symmetry meridian plane $z=0$, $x\in[-3,2]$, $y\in[-3,2]$ (red); B, transverse cut $x=-0.5$ between the duct exit and the leading edge, $y\in[-3,2]$, $z\in[-3,3]$ (blue); C, horizontal plane $y=0$ through the duct axis, $x\in[-3,2]$, $z\in[-3,3]$ (orange); D, transverse cut $x=+0.5$ downstream of the leading edge, $y\in[-3,2]$, $z\in[-3,3]$ (violet).}
  \label{fig:planes}
\end{figure}

All results presented below are directly from the theoretical model. 
The calculations are based on a canonical geometry (figure~\ref{fig:planes}): a single cylindrical duct of radius \(D=1\), without a hub (\(h=0\)), occupying \(x\le -2\) with exit centre at \((x,r)=(-2,0)\); and a rigid semi-infinite plate at \(y=-2\), occupying \(x\le 0\), giving an exit-to-edge streamwise distance \(L=2\). For the same incident spinning mode \((m,n)=(5,1)\) at \(\omega=2\pi fD/c_0=20\), two flow configurations are considered: (i) a fully quiescent configuration (\(M_0=M_1=0\)), in which no vortex sheet is shed from the lip and the Kutta parameter \(\gamma_K\) of (\ref{eq:Fp}) is immaterial, with the mode cut on at axial wavenumber \(\mu=0.947\); and (ii) a jet configuration, with \(M_1=0.353\), density ratio unity, and the full Kutta condition \(\gamma_K=1\) at the lip, for which \(\mu=0.686\). 

The quiescent case is included deliberately as a purely acoustic benchmark, facilitating direct validation against numerical simulations; the jet case, by contrast, exercises the full coupled model. A third configuration---\((m,n)=(13,1)\), \(\omega=28.3\), \(M_1=115/340=0.338\)---is reserved for the far-field study. The near field is examined in the four cut planes defined in figure~\ref{fig:planes}: the symmetry meridian plane \(z=0\) (A, \(x\in[-3,2]\), \(y\in[-3,2]\)); the transverse cut \(x=-0.5\) between the duct exit and the edge (B, \(y\in[-3,2]\), \(z\in[-3,3]\)); the horizontal plane \(y=0\) through the duct axis (C, \(x\in[-3,2]\), \(z\in[-3,3]\)); and the transverse cut \(x=+0.5\) downstream of the edge (D, \(y\in[-3,2]\), \(z\in[-3,3]\)).

As a comparison, the free-field radiation $p_d$ of (\ref{eq:ductfield}), radiated by the duct without the plate, is the input to the second stage; its validation against the reference duct code \citep{huang2008} is reported in \S\ref{sec:validation}.

\begin{figure}
  \centering
  \subfigure[]{\includegraphics[width=0.49\textwidth]{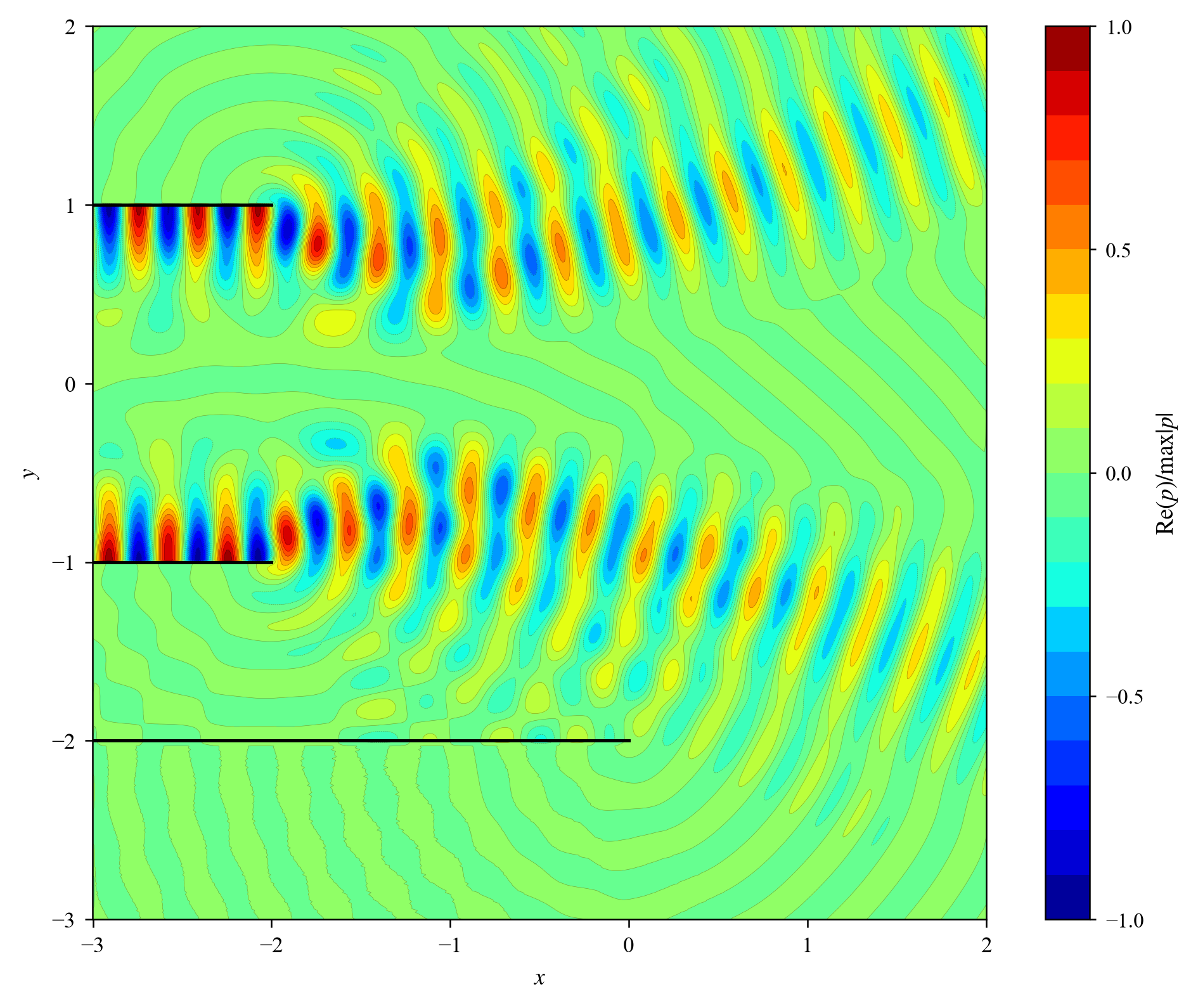}}\hfill
  \subfigure[]{\includegraphics[width=0.49\textwidth]{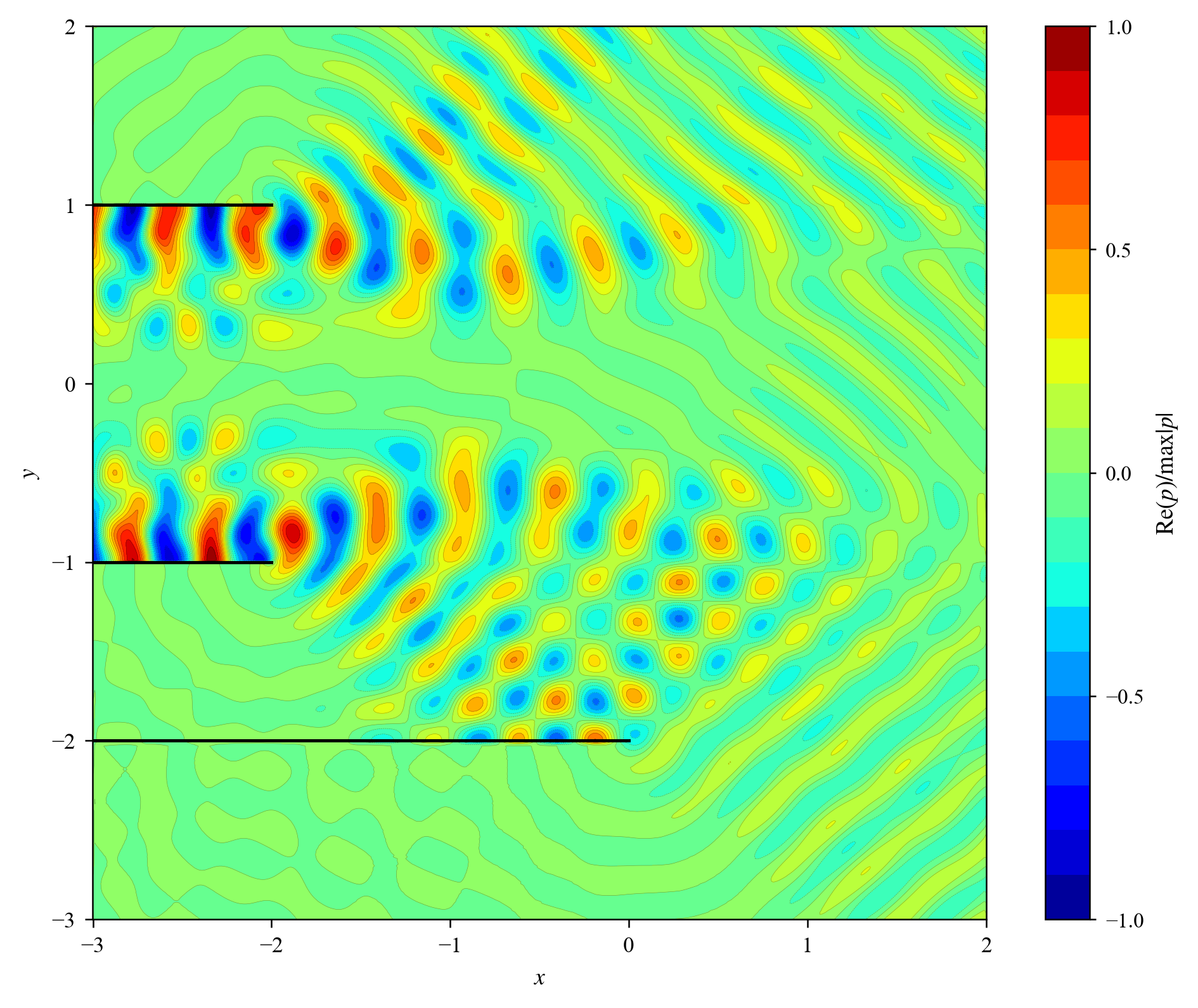}}
  \caption{Instantaneous total pressure  in the symmetry meridian plane $z=0$ (plane A), mode $(m,n)=(5,1)$, $\omega=20$: (a) quiescent configuration ($M_0=M_1=0$); and (b) with-jet configuration ($M_1=0.353$, $M_0=0$). Each panel is normalised by its own maximum absolute value on the plane, with 21 equally spaced contour levels in $[-1,1]$. The duct wall ($|y|\le1$, $x\le-2$) and the semi-infinite plate ($y=-2$, $x\le0$) are outlined.}
  \label{fig:planeA}
\end{figure}

Figure~\ref{fig:planeA} shows the instantaneous total pressure in the symmetry meridian plane $z=0$ (plane A) for the two flow configurations. The duct radiation pattern, with the conical beam structure typical of a spinning mode, is clearly seen in both; the plate lies along $y=-2$ for $x\le0$. Two differences stand out. First, the axial wavelength inside the duct is visibly longer in the with-jet case, reflecting the smaller axial wavenumber parameter ($\mu=0.686$ against $0.947$). Secondly, the radiated beams are refracted away from the downstream direction by the jet: the wavefronts outside the duct lean markedly more steeply in Figure~\ref{fig:planeA}(b), corresponding to a far-field main lobe at $\varphi\approx48^\circ$ against $\varphi\approx22^\circ$ in the quiescent case. In both configurations the incident radiation interferes with the field scattered by the plate, producing the standing-wave pattern visible between the duct and the plate, while below the plate ($y<-2$) only edge diffraction penetrates the geometrical shadow. The chirality of the spinning mode is not visible in this plane (which contains the duct axis); it appears in the transverse cuts B and D below.

\begin{figure}
  \centering
  \subfigure[]{\includegraphics[width=0.49\textwidth]{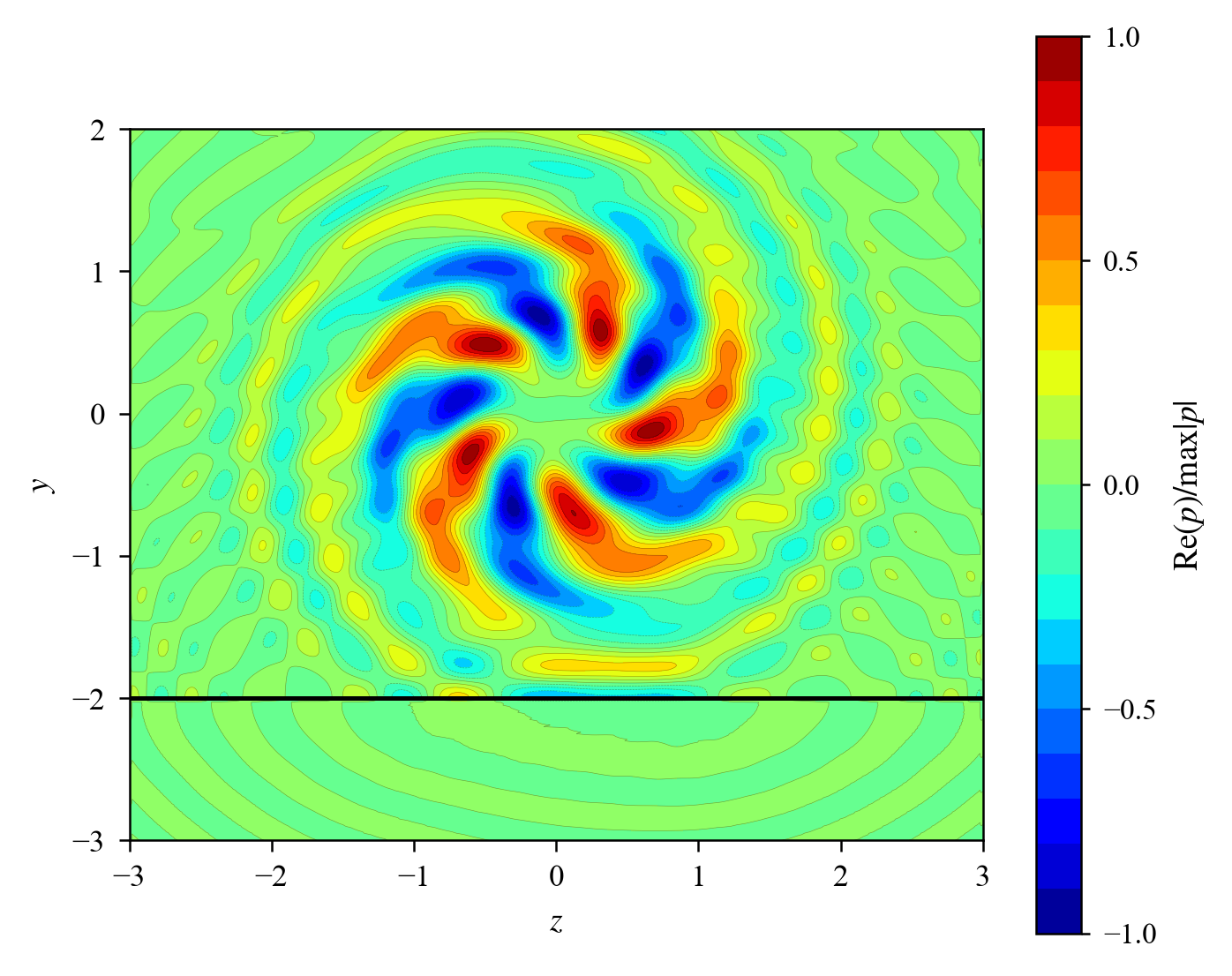}}\hfill
  \subfigure[]{\includegraphics[width=0.49\textwidth]{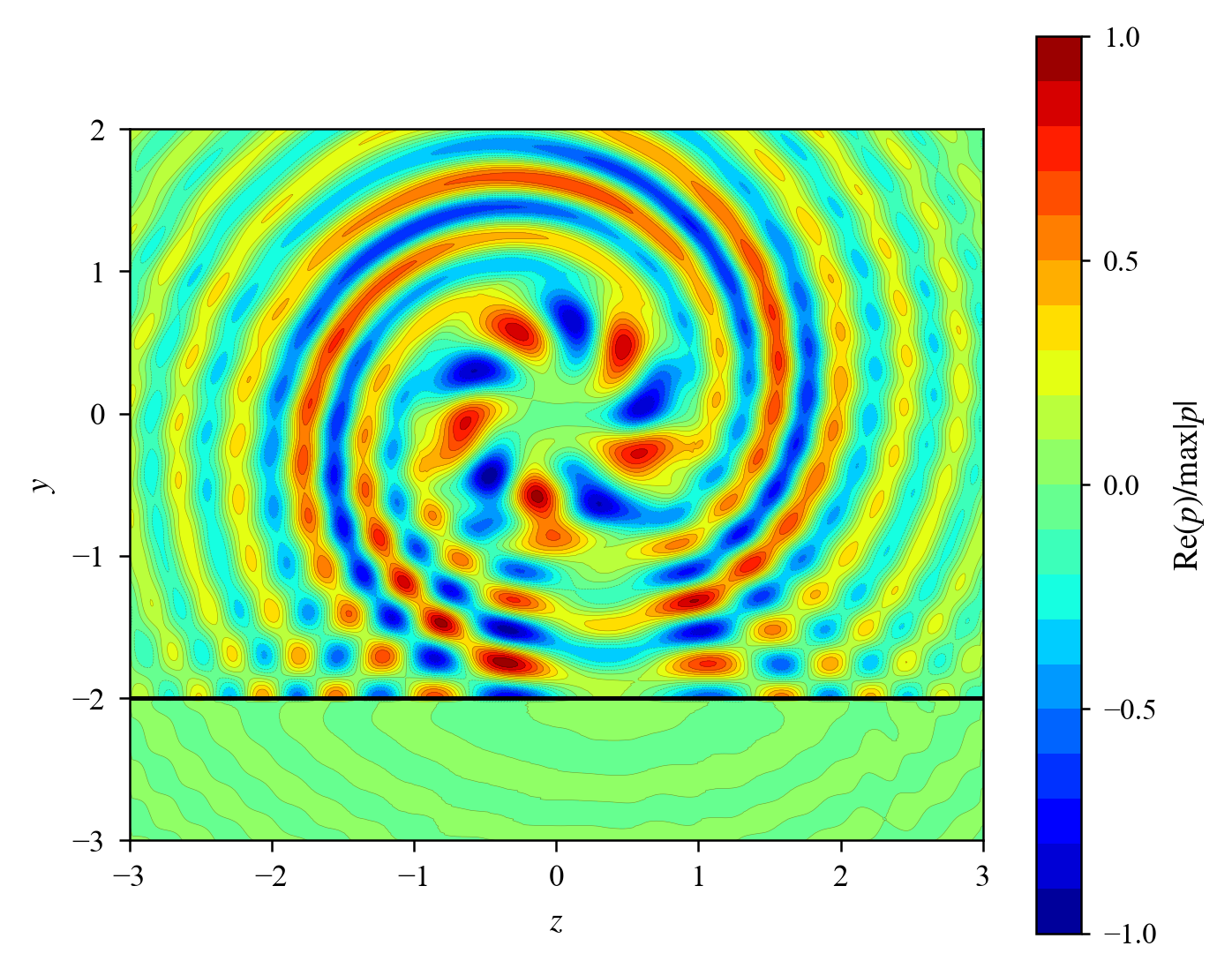}}
  \caption{Instantaneous total pressure $\mathrm{Re}(p)$ in the transverse cut $x=-0.5$ between the duct exit and the plate edge (plane B), shown in the $(z,y)$ plane: (a) quiescent; (b) with jet. Normalisation, contour levels and colour scale as in figure~\ref{fig:planeA}. The plate at $y=-2$ is outlined.}
  \label{fig:planeB}
\end{figure}

Figure~\ref{fig:planeB} shows the total field in the transverse cut $x=-0.5$ between the duct exit and the leading edge (plane B). The $m=5$ spiral structure of the spinning mode around the duct cross-section (with its $2m$-lobed azimuthal modulation), the interference between the direct and edge-diffracted fields outside the spiral, and the shadow zone below the plate ($y<-2$) are apparent in both configurations. The with-jet result of figure~\ref{fig:planeB}(b) shows a denser ring structure around the spiral arms, reflecting the shorter radial wavelength of the radiated field in the presence of the jet.

\begin{figure}
  \centering
  \subfigure[]{\includegraphics[width=0.49\textwidth]{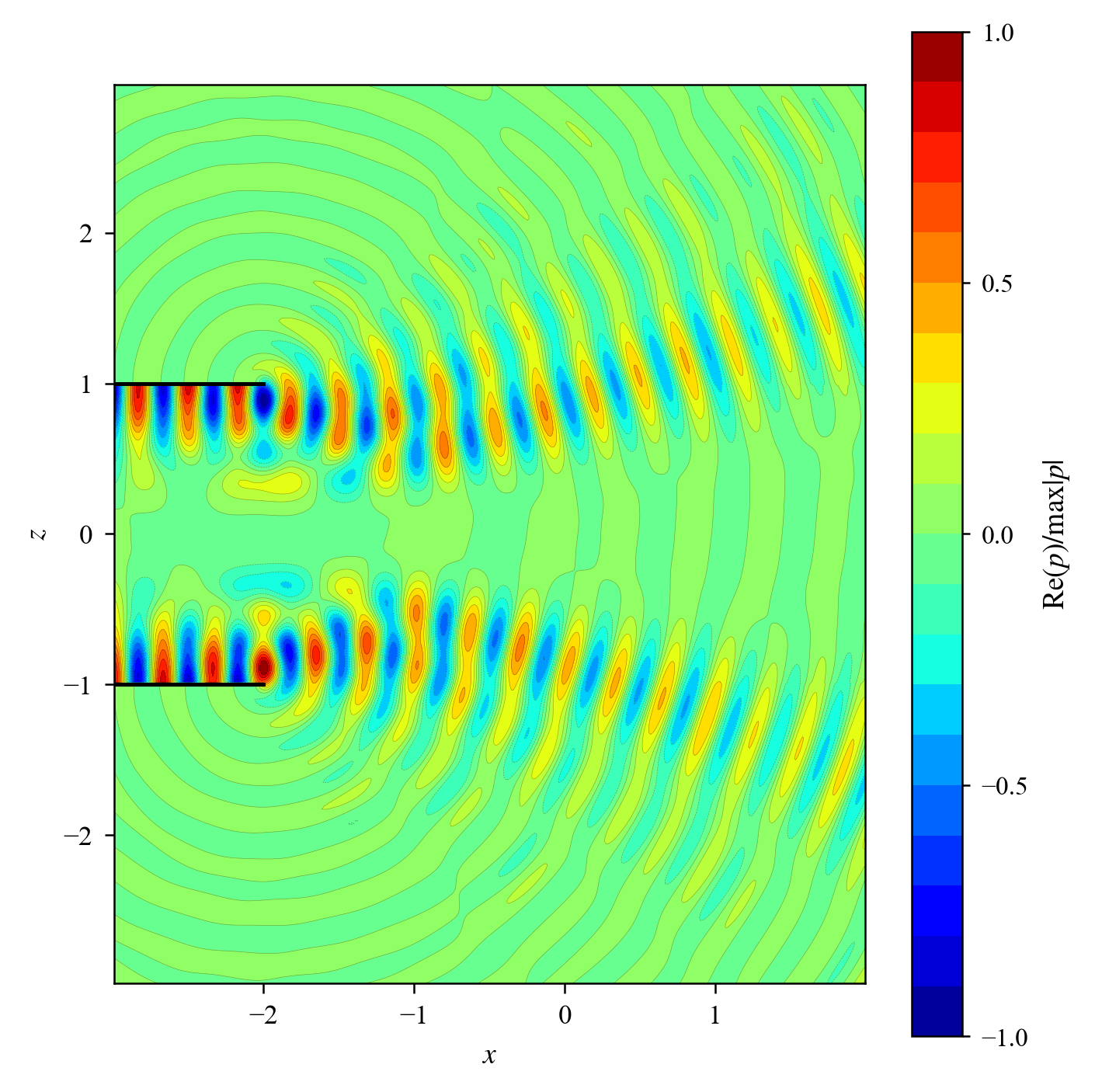}}\hfill
  \subfigure[]{\includegraphics[width=0.49\textwidth]{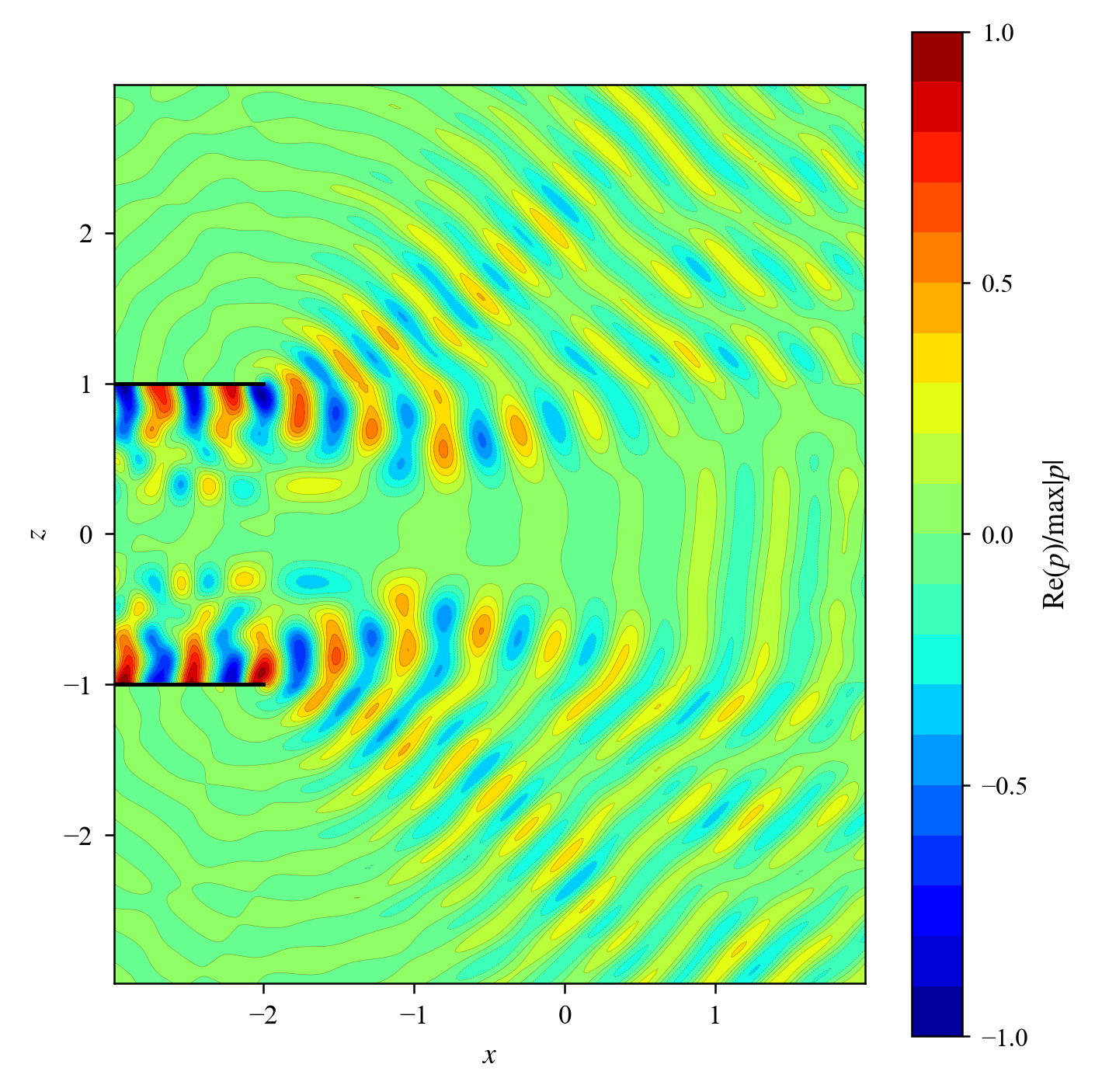}}
  \caption{Instantaneous total pressure $\mathrm{Re}(p)$ in the horizontal plane $y=0$ through the duct axis (plane C), shown in the $(x,z)$ plane: (a) quiescent; (b) with jet. Normalisation, contour levels and colour scale as in figure~\ref{fig:planeA}. The duct wall generators $z=\pm1$ ($x\le-2$) are outlined.}
  \label{fig:planeC}
\end{figure}

Figure~\ref{fig:planeC} shows the total field in the horizontal plane $y=0$ through the duct axis (plane C). This plane contains the duct axis and cuts the duct wall along its generators $z=\pm1$; comparing with figure~\ref{fig:planeA} exposes the lateral (spanwise) influence of the plate on the radiated field. The longer axial wavelength and the steeper radiation angle of the with-jet case are again visible in figure~\ref{fig:planeC}(b).

\begin{figure}
  \centering
  \subfigure[]{\includegraphics[width=0.49\textwidth]{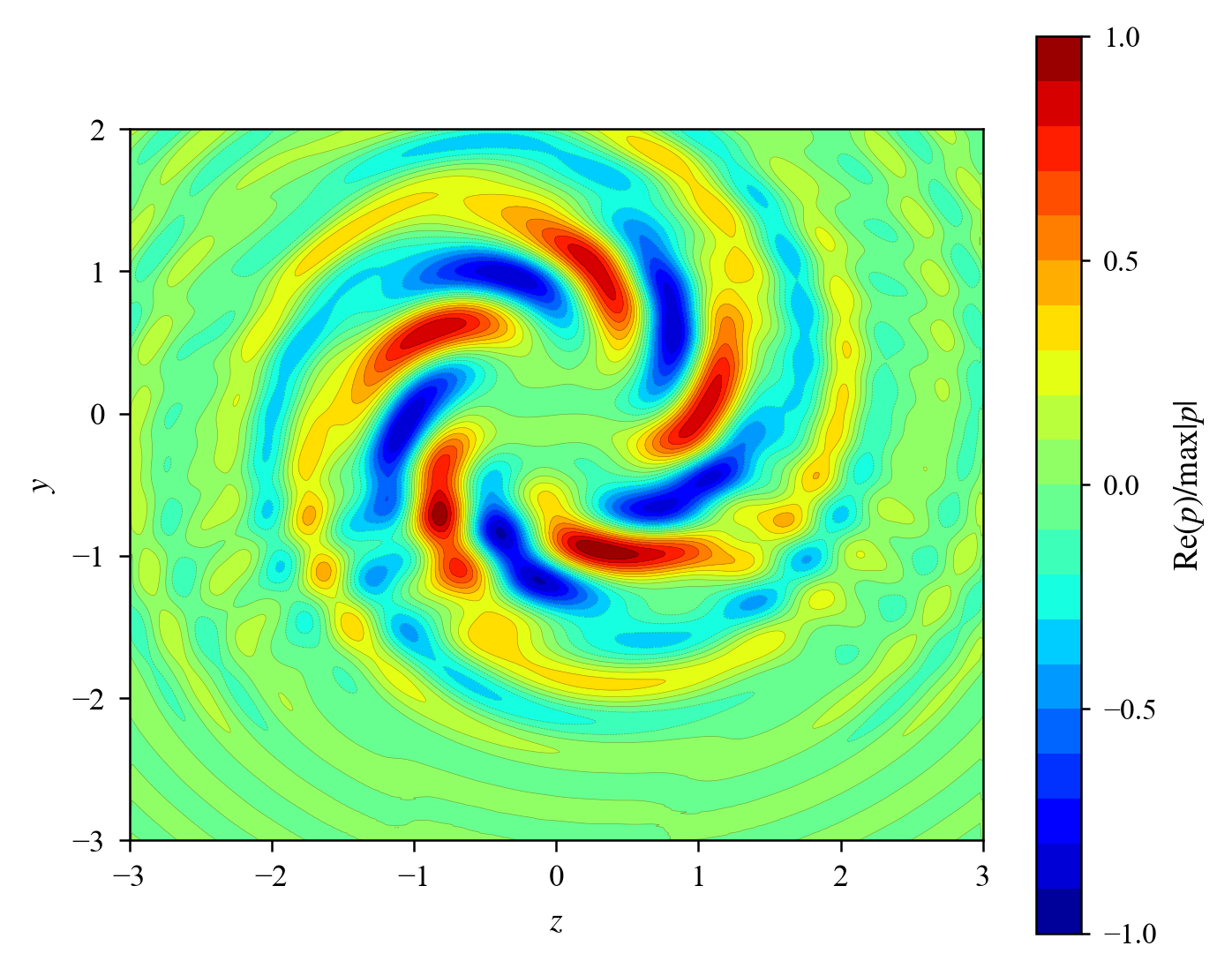}}\hfill
  \subfigure[]{\includegraphics[width=0.49\textwidth]{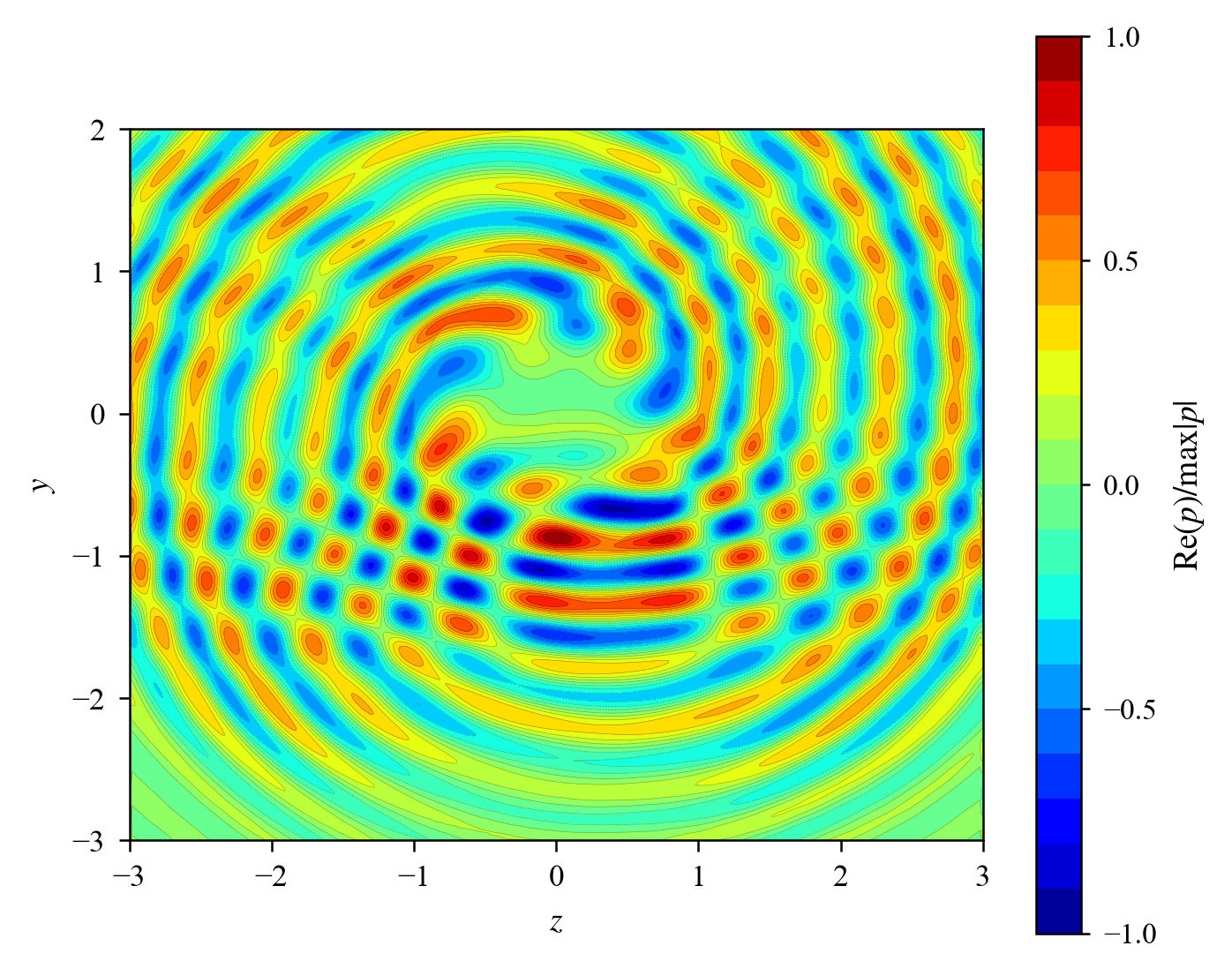}}
  \caption{Instantaneous total pressure $\mathrm{Re}(p)$ in the transverse cut $x=+0.5$ just downstream of the plate edge (plane D), shown in the $(z,y)$ plane: (a) quiescent; (b) with jet. Normalisation, contour levels and colour scale as in figure~\ref{fig:planeA}. The plate does not extend into this cut; the region $y<-2$ lies in its geometrical shadow.}
  \label{fig:planeD}
\end{figure}

Figure~\ref{fig:planeD} shows the total field in the transverse cut $x=+0.5$ just downstream of the leading edge (plane D), where the edge-diffracted field emanating from the edge line interferes with the direct radiation above the plate and the shadow zone below it ($y<-2$). In both configurations the spiral of the incident mode is cut across at the plate level and the shadow below is partly filled by edge diffraction; the with-jet result of figure~\ref{fig:planeD}(b) shows in addition a pronounced horizontal fringe pattern just above the plate level, produced by interference between the direct field and the field returned from the plate.

A remark on symmetry is in order before turning to the far field. The near field results in figures~\ref{fig:planeB}--\ref{fig:planeD} are not symmetric about $z=0$. The reason is the chirality (in $\theta$ direction) of the incident spinning mode: the azimuthal dependence $\mathrm{exp}({\ii m\theta})$ is mapped to $\mathrm{exp}({-\ii m\theta})$ by the mirror operation $z\to-z$ (which sends $\theta\to-\theta$), so a single travelling spinning mode is never mirror-symmetric -- only $m=0$ or an equal superposition of $+m$ and $-m$ (a standing azimuthal wave) is. Within the present model the asymmetry enters explicitly through the Weyl--Sommerfeld algebraic factor $((\gamma_d-\ii\nu)/k)^m$ of (\ref{eq:Ahat}), which is not an even function of $\nu$ for $m\neq0$: the spanwise spectrum of the incident field is shifted and shaped according to the sign of $m$, and the plate scatters this asymmetric spectrum accordingly. The same mechanism is visible directly on the plate: the incident normal velocity $v_i=\partial_y\psi_d$ contains, beside a term even in $z$, the azimuthal-derivative contribution $-\ii m(z/r_p^2)\,\psi_d$, which is odd in $z$. 

\begin{figure}
  \centering
  \subfigure[]{\includegraphics[width=0.5\textwidth]{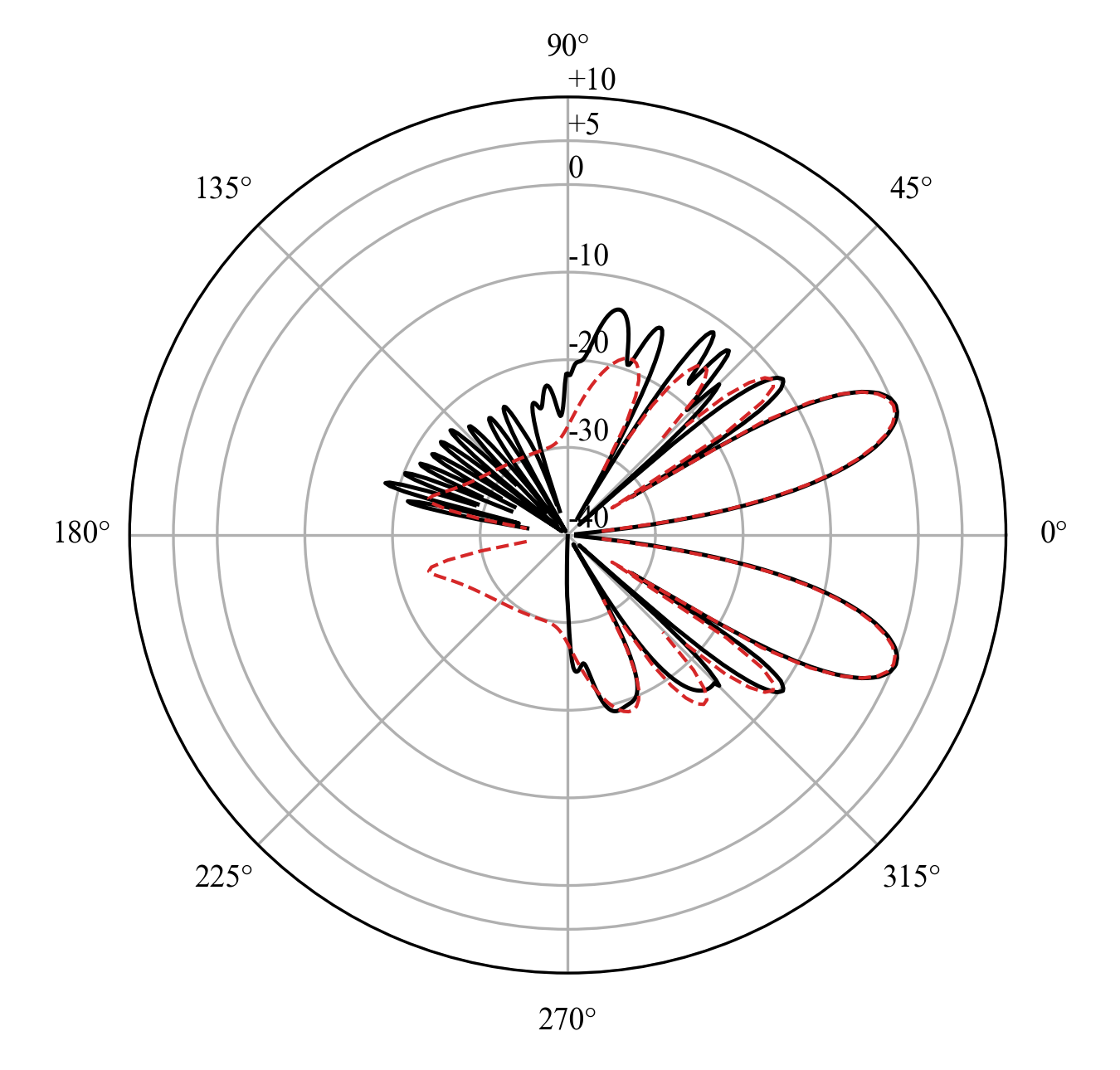}}\hfill
  \subfigure[]{\includegraphics[width=0.5\textwidth]{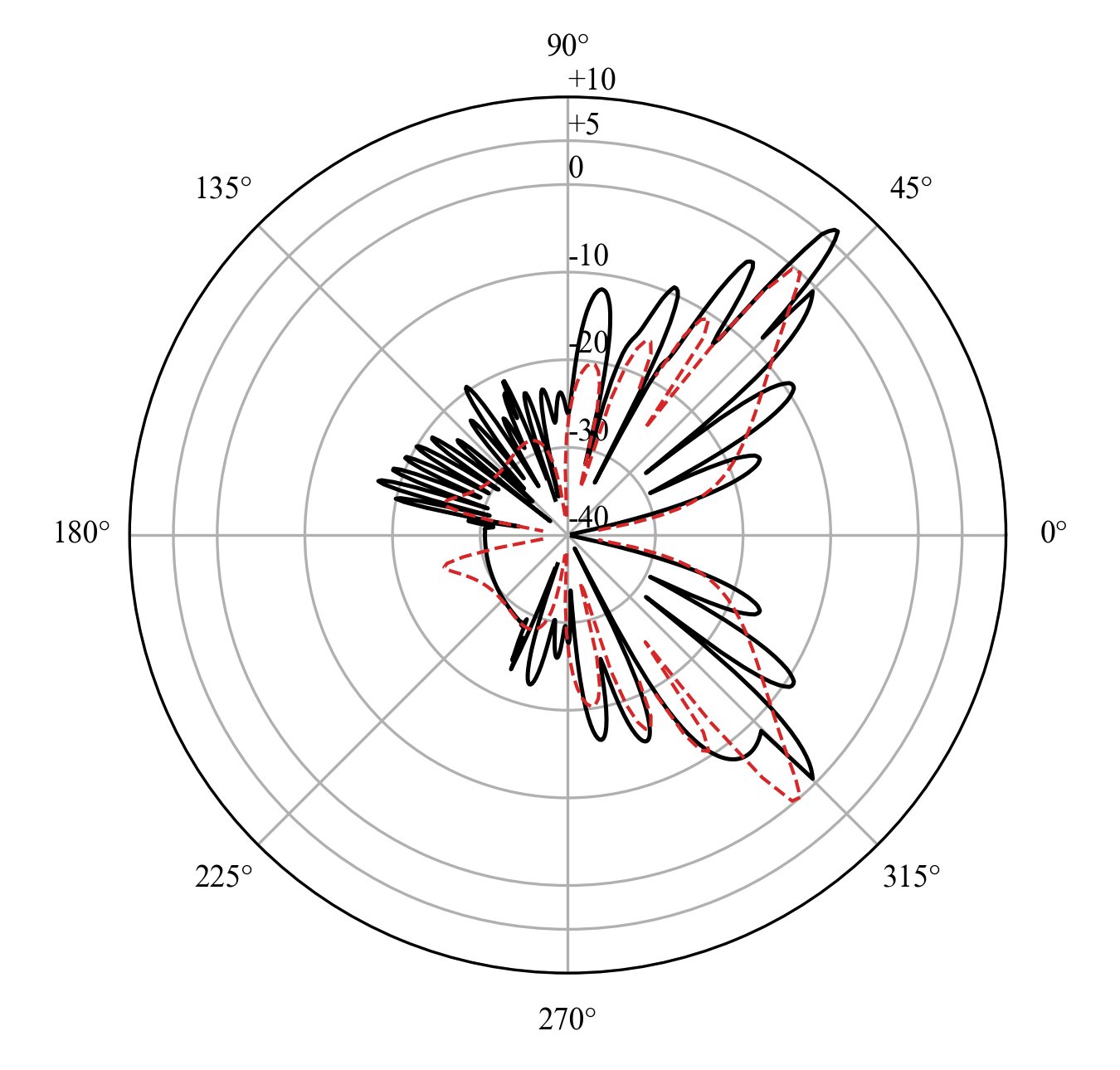}}\hfill
  \subfigure[]{\includegraphics[width=0.5\textwidth]{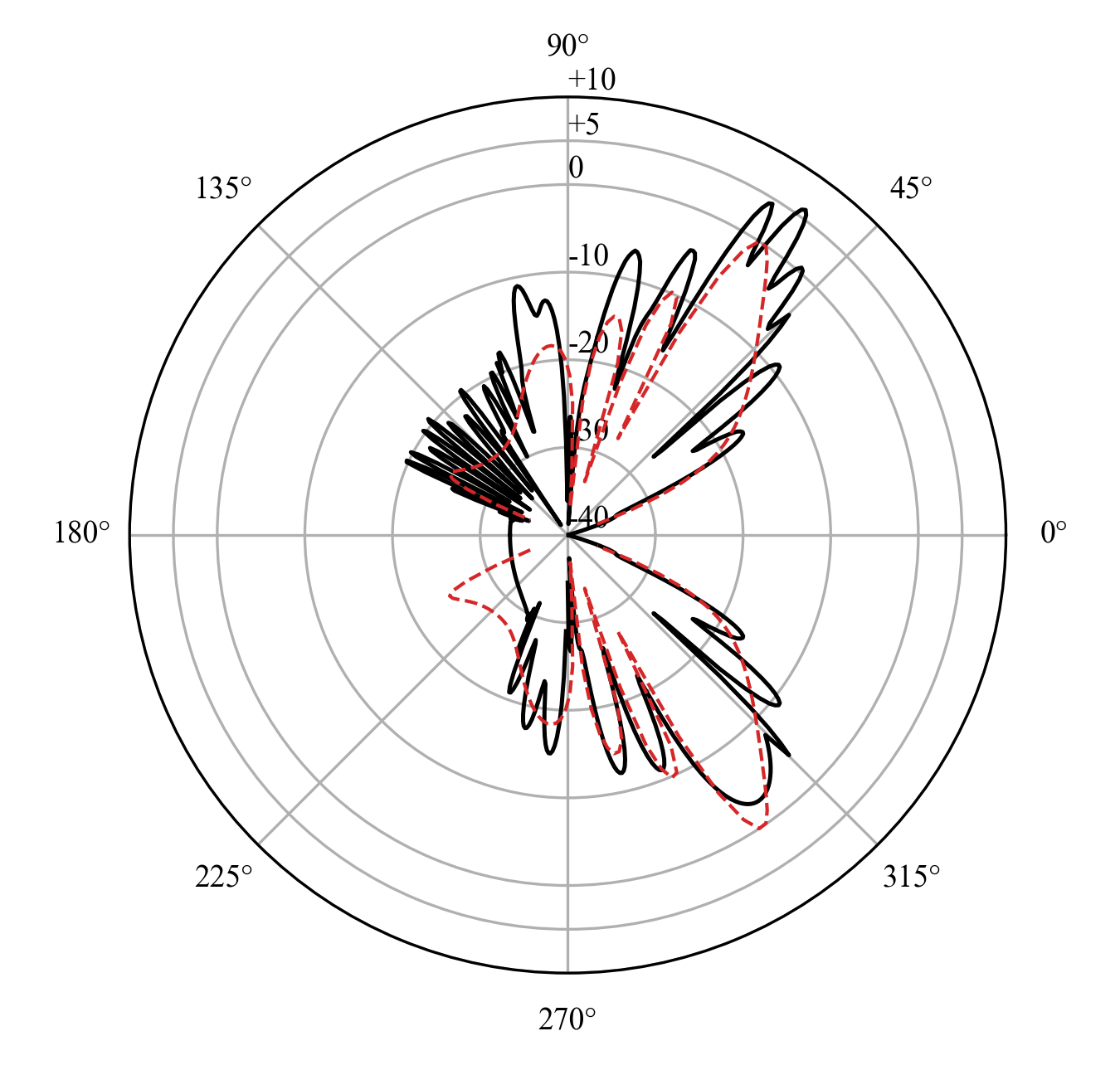}}
  \caption{Full-circle far-field directivity $20\log_{10}(|\mathcal D_{\rm tot}|/\max|D_P|)$ in the $z=0$ plane: the coupled solution (\ref{eq:Dtot}) with the plate (black solid) and the duct-alone directivity $|D_P|$ of (\ref{eq:ductdir}) (red dashed), both normalised by the duct-alone maximum; radial range $-40$ to $+10$~dB; polar angle $\varphi$ measured from the $+x$ direction about the plate edge. (a) Quiescent configuration, $(m,n)=(5,1)$, $\omega=20$, $M_0=M_1=0$; (b) the with-jet configuration, $(m,n)=(5,1)$, $\omega=20$, $M_1=0.353$; and (c) the higher-order case, $(m,n)=(13,1)$, $\omega=28.3$, $M_1=0.338$. }  
  \label{fig:directivity} 
\end{figure}

The full-circle far-field directivity (\ref{eq:Dtot}) in the $z=0$ plane is shown in figure~\ref{fig:directivity} for the three configurations: the quiescent case (figure~\ref{fig:directivity}a), the with-jet case of the same mode (figure~\ref{fig:directivity}b), and a higher-order, higher-frequency case -- mode $(m,n)=(13,1)$, cut-on with axial wavenumber parameter $\mu=0.598$, $\omega=28.3$, jet Mach $M_1=115/340=0.338$ (figure~\ref{fig:directivity}c) -- computed to exercise the model at a richer modal content. 

In all cases the incident mode carries the unit-energy-flux normalisation of appendix~B of \citet{gabard2006} used in (\ref{eq:incmode}); since the model is linear end to end, every computed field and directivity simply scales with the incident-mode amplitude, and the normalised decibel presentation of figure~\ref{fig:directivity} is independent of this choice. An elementary check is provided by the duct-alone curves, which are symmetric about the $0^\circ$--$180^\circ$ axis, as they must be in the absence of the plate (the azimuthal factor satisfies $|\ee^{\ii m\hat\theta}|=1$), whereas the with-plate curves are strongly asymmetric. 

Three plate effects are visible, matching the three terms of (\ref{eq:Dtot}). First, in the specular zone above the plate the image-source reflection interferes with the direct radiation on a fringe pattern set by the source--image path difference $2d\sin\varphi$ (an interference phase $2\omega d\sin\varphi=80\sin\varphi$ at $\omega=20$): the duct-alone main lobe peaks at $\varphi\approx22^\circ$ (quiescent), $\varphi\approx48^\circ$ (with jet) and $\varphi\approx57^\circ$ (higher-order case), and whether main lobe is reinforced or cancelled depends on the fringe on which it falls -- the quiescent lobe sits on a nearly destructive fringe and is essentially unchanged ($+0.3$~dB), while the with-jet lobe falls on a constructive fringe and is reinforced by $6.4$~dB (at $\varphi\approx48^\circ$) and the higher-order lobe by $7.4$~dB (at $\varphi\approx54^\circ$); the deep duct-alone nulls of the specular zone are filled by up to $15$~dB (quiescent case, at $\varphi\approx45^\circ$). The jet therefore affects the installation effect twice: it refracts the main lobe away from the downstream axis (from $22^\circ$ to $48^\circ$ between figure~\ref{fig:directivity}a and figure~\ref{fig:directivity}b), and thereby moves the lobe onto a different interference fringe. 

Second, the plate casts a shadow below itself: over the downward directions $\varphi\in(200^\circ,340^\circ)$ the median level falls to $-35.9$~dB (quiescent), $-25.8$~dB (with jet) and $-23.2$~dB (higher order) relative to the duct-alone maximum, the residual field being pure edge diffraction; the level recovers towards grazing downstream directions ($\varphi\to360^\circ$), which the plate does not block. Third, the edge diffraction fills all directions, including the shadow, at a level some $20$--$40$~dB below the main lobe. The higher-order mode produces, as expected, a denser multi-lobed pattern. The isolated direction $\varphi=180^\circ$ (observer looking along the plate surface) is degenerate for the saddle formula (\ref{eq:Dpsi}) and is excluded from the plots.

A remark on the reliability of the solution in the third quadrant ($180^\circ<\varphi<270^\circ$, the deep shadow below the plate on the upstream side) is also necessary. There the directivity consists of edge diffraction alone, lying $15$--$42$~dB below the duct-alone maximum, and two limitations -- one theoretical, one numerical -- should be noted. On the theoretical side, the saddle-point assembly (\ref{eq:Dtot}) is non-uniform at the boundaries of the shadow: as $\varphi\to180^\circ$ the observation direction becomes tangent to the semi-infinite plate and the saddle $\alpha_s=-\khat n_x$ coalesces with the branch point at $\alpha=\khat$, so the ordinary stationary-phase result degenerates; likewise the exact shadow cancellation (\ref{eq:shadow}) imposes a sharp transition at the shadow boundary $\varphi\approx315^\circ$, where a uniform (Fresnel-type) treatment of the pole crossing the saddle would be needed for pointwise accuracy. These are limitations of the saddle asymptotics, not of the Wiener--Hopf solution itself, which remains exact there. On the numerical side, a direct numerical evaluation of the full spectral integral at finite radius ($R\approx20$) is itself limited by phase undersampling of the spectral grid ($\Delta\alpha\,R\sim2$) and by the non-uniformity of the asymptotics at finite $R$. Nevertheless, since the shadow-zone levels lie more than $15$~dB below the main lobe, these limitations do not affect any of the conclusions.

\section{Conclusions}
\label{sec:conclusions}

A theoretical model has been constructed for the radiation of a spinning duct mode from a semi-infinite circular duct and its scattering by a nearby semi-infinite rigid plate, by chaining two Wiener--Hopf problems through an exact spectral bridge. The first stage is the established duct-radiation Wiener--Hopf solution \citep{munt1977,rienstra1984,gabard2006}, used as a black box through its output $F_+(u)$. The bridge re-expands the incident field on the plate into a double angular spectrum of plane waves, the spanwise part being generated by a Weyl--Sommerfeld identity in which the azimuthal order $m$ enters through the algebraic factor $((\gamma-\ii\nu)/k)^m$ -- the explicit carrier of the coordinate change from the axisymmetric duct frame to the Cartesian plate frame. Each spectral component is then scattered by a scalar Wiener--Hopf problem whose kernel splits in closed form, so that the second stage is exact and numerically cheap; two saddle-point evaluations reduce the far field to a single contour integral, and the geometrical reflection is summed exactly by an image-source identity, and the full-circle directivity is assembled from the edge diffraction, the incident field with its exact shadow cancellation, and the image reflection (\ref{eq:Dtot}). 

With the theoretical model, three configurations are computed -- the mode $(m,n)=(5,1)$ at $\omega=20$, both fully quiescent and with a jet of Mach $0.353$, and the higher-order mode $(13,1)$ at $\omega=28.3$ with a jet of Mach $0.338$. In all three the plate reorganises the radiated field through image-source interference: a main lobe falling on a constructive interference fringe is reinforced by $6$--$7$~dB, the deep duct-alone nulls of the specular zone are filled by up to $15$~dB, and below itself the plate casts a shadow $23$--$36$~dB deep, filled only by edge diffraction.

The main assumptions and limitations of the theoretical model are as follows. (i) One-way coupling: the re-radiation of the scattered field back into the duct, and its subsequent re-scattering by the plate, are neglected; this is reasonable when the plate is not too close to the exit, and can be restored, if needed, by iterating the two stages. (ii) All computed configurations have a quiescent ambient ($M_0=0$); nevertheless, a uniform external flow can be accommodated by the Lorentz transformation of \citet{gabard2006} at the duct and the Prandtl--Glauert transformation of \citet{huang2017} at the plate, without any new ingredient. (iii) The jet of the first stage enters the coupling only through the exit spectrum $F_+(u)$, while the second stage treats the medium around the plate as quiescent and uniform, i.e.\ the downstream vortex sheet is not advected over the plate -- consistent with the one-way coupling and reasonable while the plate remains outside the jet ($d>1$). (iv) The Kelvin--Helmholtz instability wave is omitted from the coupling, as discussed in detail below. (v) The plate is semi-infinite, rigid and of vanishing thickness; a finite-thickness, finite-span or impedance plate would require a different stage-two kernel but the same bridge. The framework also extends, and most significantly, to a spanwise-periodic serrated edge $\chi(z)$, which promotes the scalar stage-two Wiener--Hopf equation to the Toeplitz matrix equation of \citet{huang2017}: the incident spanwise spectrum $\widehat A(u,\nu)$, whose centroid is set by the spinning order $m$, is then redistributed among the serration orders $\nu+2\piup\kappa/\lambda_s$, and orders for which $|\khat_\kappa|$ becomes imaginary cannot reach the far field. The resulting prediction -- how serrations modify the installation scattering of a spinning duct mode -- is new and shall consistitutes the subject of ongoing work.

The omission of the instability wave, listed among the assumptions above, deserves a fuller discussion because the stage-one solution formally contains it: as noted in \S\ref{sec:stage1field}, the residue of the pole at $u_0$ in (\ref{eq:ductfieldfull}) is the convective Kelvin--Helmholtz instability of the vortex sheet shed from the duct lip. This hydrodynamic disturbance grows exponentially in the streamwise direction but, for a subsonic jet, decays exponentially away from the sheet; the instability field is therefore confined to a double-conical zone about the vortex sheet \citep{gabard2006}. Its radiation mechanism is distinct from the modal radiation studied here. It should also be recalled that the unlimited exponential growth of the instability wave far downstream is an artefact of the linear theory; in a real jet the growth is rapidly saturated by nonlinear effects. 

Finally, we return to the LLM-assisted nature of this work in light of the \citet{leiden2026declaration}. Our experience suggests that the distinction between \lq\lq human idea\rq\rq and \lq\lq machine intelligence\rq\rq is not merely rhetorical but practically meaningful: the core mathematical step of this paper--the recognition that a Weyl--Sommerfeld identity can serve as an exact spectral bridge between cylindrical and Cartesian representations--was supplied by our own conception, while the LLM handled the heavy algebra of translating between the two coordinate systems and the associated numerical verification. We do not claim that LLMs are ready for unsupervised complex-variable reasoning--the kernel factorisations and causality arguments in this work remain human-derived--but we do suggest that, used as a mechanical assistant under systematic verification, an LLM can meaningfully accelerate theoretical development in applied mathematics and fluid mechanics. This case study, we hope, contributes a concrete example to the ongoing debate on AI's role in rigorous research.


\section*{Declaration of Interests}
The authors report no conflict of interest. 

\section*{Acknowledgements}
This work is partly supported by National Science Foundation of China (Grant No. 12432016 \& No. 12272007). 
The computational code developed for this work is available from the authors upon reasonable request. The algebraic bridging between the two coordinate systems and the computational pipeline were assisted by a large language model coding assistant (Kimi, Moonshot AI). 
%


\begin{thebibliography}{32}
\expandafter\ifx\csname natexlab\endcsname\relax\def\natexlab#1{#1}\fi
\def\au#1{#1} \def\ed#1{#1} \def\yr#1{#1}\def\at#1{#1}\def\jt#1{\textit{#1}}
  \def\bt#1{#1}\def\bvol#1{\textbf{#1}} \def\vol#1{#1} \def\pg#1{#1}
  \def\publ#1{#1}\def\arxiv#1{#1}\def\org#1{#1}\def\st#1{\textit{#1}}

\bibitem[Abrahams {\em et~al.\/}(2021)Abrahams, Huang, Kisil, Mishuris, Nieves,
  Rogosin \& Spitkovsky]{Huang:21nsr}
{\sc \au{Abrahams, I.~D.}, \au{Huang, X.}, \au{Kisil, A.}, \au{Mishuris, G.},
  \au{Nieves, M.}, \au{Rogosin, S.} \& \au{Spitkovsky, I.}} \yr{2021}
  \at{Reinvigorating the Wiener--Hopf technique in the pursuit of understanding
  processes and materials}.  \jt{Nat. Sci. Rev.}  \bvol{8}~(2),  \pg{nwaa225}.

\bibitem[Ayton \& Peake(2013)]{ayton2013}
{\sc \au{Ayton, L.~J.} \& \au{Peake, N.}} \yr{2013}  \at{On high-frequency
  noise scattering by aerofoils in flow}.  \jt{J. Fluid Mech.}  \bvol{734},
  \pg{144--182}.

\bibitem[Baddoo \& Ayton(2020)]{baddoo2020}
{\sc \au{Baddoo, P.~J.} \& \au{Ayton, L.~J.}} \yr{2020}  \at{Acoustic
  scattering by cascades with complex boundary conditions: compliance, porosity
  and impedance}.  \jt{J. Fluid Mech.}  \bvol{898},  \pg{A16}.

\bibitem[Cargill(1982)]{cargill1982}
{\sc \au{Cargill, A.~M.}} \yr{1982}  \at{Low-frequency sound radiation and
  generation due to the interaction of unsteady flow with a jet pipe}.  \jt{J.
  Fluid Mech.}  \bvol{121},  \pg{59--105}.

\bibitem[Crighton {\em et~al.\/}(1992)Crighton, Dowling, Ffowcs~Williams, Heckl
  \& Leppington]{crighton1992}
{\sc \au{Crighton, D.~G.}, \au{Dowling, A.~P.}, \au{Ffowcs~Williams, J.~E.},
  \au{Heckl, M.} \& \au{Leppington, F.~G.}} \yr{1992} {\em Modern Methods in
  Analytical Acoustics\/}.  \publ{London: Springer-Verlag}.

\bibitem[Ffowcs~Williams \& Hall(1970)]{ffowcs1970}
{\sc \au{Ffowcs~Williams, J.~E.} \& \au{Hall, L.~H.}} \yr{1970}
  \at{Aerodynamic sound generation by turbulent flow in the vicinity of a
  scattering half plane}.  \jt{J. Fluid Mech.}  \bvol{40}~(4),  \pg{657--670}.

\bibitem[Gabard \& Astley(2006)]{gabard2006}
{\sc \au{Gabard, G.} \& \au{Astley, R.~J.}} \yr{2006}  \at{Theoretical model
  for sound radiation from annular jet pipes: far- and near-field solutions}.
  \jt{J. Fluid Mech.}  \bvol{549},  \pg{315--341}.

\bibitem[Hales {\em et~al.\/}(2023)Hales, Ayton, Jiang, Mahgoub, Kisler, Dixon,
  de~Silva, Moreau \& Doolan]{hales2023}
{\sc \au{Hales, A. D.~G.}, \au{Ayton, L.~J.}, \au{Jiang, C.}, \au{Mahgoub, A.},
  \au{Kisler, R.}, \au{Dixon, R.}, \au{de~Silva, C.}, \au{Moreau, D.} \&
  \au{Doolan, C.}} \yr{2023}  \at{A mathematical model for the interaction of
  anisotropic turbulence with a rigid leading edge}.  \jt{J. Fluid Mech.}
  \bvol{970},  \pg{A29}.

\bibitem[Howe(1991)]{howe1991}
{\sc \au{Howe, M.~S.}} \yr{1991}  \at{Aerodynamic noise of a serrated trailing
  edge}.  \jt{J. Fluids Struct.}  \bvol{5}~(1),  \pg{33--45}.

\bibitem[Huang(2017)]{huang2017}
{\sc \au{Huang, X.}} \yr{2017}  \at{Theoretical model of acoustic scattering
  from a flat plate with serrations}.  \jt{J. Fluid Mech.}  \bvol{819},
  \pg{228--257}.

\bibitem[Huang {\em et~al.\/}(2008)Huang, Chen, K. \& Zhang]{huang2008}
{\sc \au{Huang, X.}, \au{Chen, X.~X.}, \au{K., Ma~Z.} \& \au{Zhang, X.}}
  \yr{2008}  \at{Efficient computation of spinning modal radiation through an
  engine bypass duct}.  \jt{AIAA J.}  \bvol{46}~(2),  \pg{1413--1422}.

\bibitem[Jing \& Du(2026)]{jing2026}
{\sc \au{Jing, X.} \& \au{Du, Z.}} \yr{2026}  \at{Edge diffraction of in-duct
  modal waves by unflanged or flanged circular duct terminations}.  \jt{J.
  Fluid Mech.}  \bvol{1034},  \pg{A33}.

\bibitem[{Leiden Declaration}(2026)]{leiden2026declaration}
{\sc \au{{Leiden Declaration}}} \yr{2026} Leiden declaration on artificial
  intelligence and mathematics. Endorsed by the International Mathematical
  Union.

\bibitem[Levine \& Schwinger(1948)]{levine1948}
{\sc \au{Levine, H.} \& \au{Schwinger, J.}} \yr{1948}  \at{On the radiation of
  sound from an unflanged circular pipe}.  \jt{Phys. Rev.}  \bvol{73},
  \pg{383--406}.

\bibitem[Liu {\em et~al.\/}(2024)Liu, Shen \& Wei]{liu2024}
{\sc \au{Liu, J.}, \au{Shen, L.} \& \au{Wei, G.-W.}} \yr{2024}  \at{{ChatGPT}
  for computational topology}.  \jt{Found. Data Sci.}  \bvol{6}~(2),
  \pg{221--250}.

\bibitem[Liu {\em et~al.\/}(2016)Liu, Jiang, Huang \& Chen]{huang2016}
{\sc \au{Liu, X.}, \au{Jiang, H.~B.}, \au{Huang, X.} \& \au{Chen, S.~Y.}}
  \yr{2016}  \at{Theoretical model of scattering from flow ducts with
  semi-infinite axial liner splices}.  \jt{J. Fluid Mech.}  \bvol{786},
  \pg{62--83}.

\bibitem[Lorteau {\em et~al.\/}(2022)Lorteau, Le~Garrec, Daroukh \&
  Polacsek]{lorteau2022}
{\sc \au{Lorteau, M.}, \au{Le~Garrec, T.}, \au{Daroukh, M.} \& \au{Polacsek,
  C.}} \yr{2022} Acoustic assessment of {BLI} effects on {A}irbus {N}autilus
  engine integration concept -- {P}art {II}: noise radiation.  \bt{In {\em 28th
  AIAA/CEAS Aeroacoustics Conference\/}, AIAA 2022-2992}.

\bibitem[Lorteau {\em et~al.\/}(2019)Lorteau, Wiart, Kopiev \&
  Denisov]{lorteau2019}
{\sc \au{Lorteau, M.}, \au{Wiart, L.}, \au{Kopiev, V.} \& \au{Denisov, S.~L.}}
  \yr{2019} Numerical study, with experimental validation, of fan noise
  installation effects in over-wing nacelle configuration using the immersed
  boundary method.  \bt{In {\em 25th AIAA/CEAS Aeroacoustics Conference\/}}.

\bibitem[Meloni {\em et~al.\/}(2021)Meloni, Proen{\c c}a, Lawrence \&
  Camussi]{meloni2021}
{\sc \au{Meloni, S.}, \au{Proen{\c c}a, A.~R.}, \au{Lawrence, J. L.~T.} \&
  \au{Camussi, R.}} \yr{2021}  \at{An experimental investigation into
  model-scale installed jet--pylon--wing noise}.  \jt{J. Fluid Mech.}
  \bvol{929},  \pg{A4}.

\bibitem[Munt(1977)]{munt1977}
{\sc \au{Munt, R.~M.}} \yr{1977}  \at{The interaction of sound with a subsonic
  jet issuing from a semi-infinite cylindrical pipe}.  \jt{J. Fluid Mech.}
  \bvol{83},  \pg{609--640}.

\bibitem[Munt(1990)]{munt1990}
{\sc \au{Munt, R.~M.}} \yr{1990}  \at{Acoustic transmission properties of a jet
  pipe with subsonic jet flow: {I}. the cold jet reflection coefficient}.
  \jt{J. Sound Vib.}  \bvol{142}~(3),  \pg{413--436}.

\bibitem[Noble(1958)]{noble1958}
{\sc \au{Noble, B.}} \yr{1958} {\em Methods Based on the {Wiener--Hopf}
  Technique for the Solution of Partial Differential Equations\/}.
  \publ{London: Pergamon Press}.

\bibitem[Redonnet {\em et~al.\/}(2010)Redonnet, Desquesnes, Manoha \&
  Parzani]{redonnet2010}
{\sc \au{Redonnet, S.}, \au{Desquesnes, G.}, \au{Manoha, E.} \& \au{Parzani,
  C.}} \yr{2010}  \at{Numerical study of acoustic installation effects with a
  computational aeroacoustics method}.  \jt{AIAA J.}  \bvol{48}~(5),
  \pg{929--937}.

\bibitem[Rienstra(1981)]{rienstra1981}
{\sc \au{Rienstra, S.~W.}} \yr{1981}  \at{Sound diffraction at a trailing
  edge}.  \jt{J. Fluid Mech.}  \bvol{108},  \pg{443--460}.

\bibitem[Rienstra(1984)]{rienstra1984}
{\sc \au{Rienstra, S.~W.}} \yr{1984}  \at{Acoustic radiation from a
  semi-infinite annular duct in a uniform subsonic mean flow}.  \jt{J. Sound
  Vib.}  \bvol{94}~(2),  \pg{267--288}.

\bibitem[Rienstra(2021)]{rienstra2021}
{\sc \au{Rienstra, S.~W.}} \yr{2021}  \at{Slowly varying modes in a
  two-dimensional duct with shear flow and lined walls}.  \jt{J. Fluid Mech.}
  \bvol{906},  \pg{A23}.

\bibitem[Romera-Paredes {\em et~al.\/}(2024)Romera-Paredes, Barekatain,
  Novikov, Balog, Kumar, Dupont, Ruiz, Ellenberg, Wang, Fawzi, Kohli \&
  Fawzi]{romera2024}
{\sc \au{Romera-Paredes, B.}, \au{Barekatain, M.}, \au{Novikov, A.}, \au{Balog,
  M.}, \au{Kumar, M.~P.}, \au{Dupont, E.}, \au{Ruiz, F. J.~R.}, \au{Ellenberg,
  J.~S.}, \au{Wang, P.}, \au{Fawzi, O.}, \au{Kohli, P.} \& \au{Fawzi, A.}}
  \yr{2024}  \at{Mathematical discoveries from program search with large
  language models}.  \jt{Nature}  \bvol{625},  \pg{468--475}.

\bibitem[Sugimoto(2022)]{sugimoto2022}
{\sc \au{Sugimoto, N.}} \yr{2022}  \at{Thermoacoustic effects on the
  propagation of non-planar sound in a circular duct}.  \jt{J. Fluid Mech.}
  \bvol{945},  \pg{A26}.

\bibitem[Trinh {\em et~al.\/}(2024)Trinh, Wu, Le, He \& Luong]{trinh2024}
{\sc \au{Trinh, T.~H.}, \au{Wu, Y.}, \au{Le, Q.~V.}, \au{He, H.} \& \au{Luong,
  T.}} \yr{2024}  \at{Solving olympiad geometry without human demonstrations}.
  \jt{Nature}  \bvol{625},  \pg{476--482}.

\bibitem[Veitch \& Peake(2008)]{veitch2008}
{\sc \au{Veitch, B.} \& \au{Peake, N.}} \yr{2008}  \at{Acoustic propagation and
  scattering in the exhaust flow from coaxial cylinders}.  \jt{J. Fluid Mech.}
  \bvol{613},  \pg{275--307}.

\bibitem[Wang \& Huang(2025)]{wang2025}
{\sc \au{Wang, R.} \& \au{Huang, X.}} \yr{2025}  \at{Analytical study of sound
  radiation from semi-infinite elliptic ducts with uniform subsonic flow}.
  \jt{J. Fluid Mech.}  \bvol{1019},  \pg{A9}.

\bibitem[Watson(1966)]{watson1966}
{\sc \au{Watson, G.~N.}} \yr{1966} {\em A Treatise on the Theory of Bessel
  Functions\/}, 2nd edn.  \publ{Cambridge University Press}.

\end{thebibliography}

\end{document}